\documentclass[twocolumn, aps, pra, superscriptaddress]{revtex4-1}
\usepackage[T1]{fontenc}
\usepackage[latin9]{inputenc}
\usepackage{color}
\usepackage{hyperref}
\hypersetup{colorlinks=true,citecolor=blue,linkcolor=blue,filecolor=blue,urlcolor=blue,breaklinks=true}
\usepackage{bm}
\usepackage{amsmath}
\usepackage{amssymb}
\usepackage{mathtools}
\usepackage{graphicx}
\usepackage{enumerate}
\usepackage[sort&compress]{natbib}
\usepackage{braket}
\usepackage{psfrag}
\usepackage{tikz}
\usetikzlibrary{calligraphy}
\usetikzlibrary{decorations.pathreplacing}
\usetikzlibrary{chains}
\usetikzlibrary{fit}
\usepackage[normalem]{ulem}
\usepackage{qcircuit}
\usepackage{subcaption}
\usepackage{accents}
\usepackage{soul}
\usepackage{algorithm}
\usepackage{algpseudocode}
\usepackage{multirow}
\usepackage{makecell}
\usepackage{url}
\usepackage{xcolor}
\usepackage{colortbl}
\usepackage{environ}
\usepackage{amsthm}
\usepackage{dsfont}
\usepackage{bm}

\newtheorem{theorem}{Theorem}

\NewEnviron{es}{%
   \begin{equation}\begin{split}
     \BODY
   \end{split}\end{equation}
}

\graphicspath{{figs/}}

\DeclareMathOperator{\Var}{Var}
\DeclareMathOperator{\I}{Im}
\DeclareMathOperator{\R}{Re}
\DeclareMathOperator{\ee}{e}

\newcommand{\norm}[1]{\left\lVert#1\right\rVert}
\newcommand{\proj}[1]{|#1\rangle\!\langle #1|}

\newcommand*{\di}{\mathrm{d}} 

\begin{document}

\title{Error Bounds for Variational Quantum Time Evolution}
\author{Christa Zoufal}
\affiliation{IBM Quantum, IBM Research Europe -- Zurich}
\affiliation{Institute for Theoretical Physics, ETH Zurich}
\author{David Sutter}
\affiliation{IBM Quantum, IBM Research Europe -- Zurich}
\author{Stefan Woerner}
\affiliation{IBM Quantum, IBM Research Europe -- Zurich}

\begin{abstract}
Variational quantum time evolution allows us to simulate the time dynamics of quantum systems with near-term compatible quantum circuits. 
Due to the variational nature of this method the accuracy of the simulation is a priori unknown.
We derive global phase agnostic error bounds for the state simulation accuracy with variational quantum time evolution that improve the tightness of fidelity estimates over existing error bounds. 
These analysis tools are practically crucial for assessing the quality of the simulation and making informed choices about simulation hyper-parameters.
The efficient, a posteriori evaluation of the bounds can be tightly integrated with the variational time simulation and, hence, results in a minor resource overhead which is governed by the system's energy variance.
The performance of the novel error bounds is demonstrated on numerical examples.
\end{abstract}

\maketitle

\section{Introduction}
\textit{Quantum time evolution} (QTE) generally describes the process of evolving a quantum state over time with respect to a Hamiltonian $H$.
In quantum real time evolution (QRTE), which is BQP-complete~\cite{cmp18}, a state is evolved according to unitary quantum dynamics of the form $e^{-i H t/\hbar}$ for $t \in \mathbb{R}$, e.g., simulating many-body dynamics \cite{BarendsFermionicModels2015, SmithSimulatingQuantumManyBody2019, barison2021efficient}.
If the time parameter $t$ is replaced by an imaginary time $\tau=-i t$ the system dynamics change to a non-unitary quantum imaginary time evolution (QITE) $e^{- H \tau/\hbar}$ which is believed to be QMA-hard~\cite{kkr06}. Hence, even a quantum computer is not expected to enable an efficient and exact execution of a generic form of these dynamics but might only perform well for certain instances. Finding these instances, or approximations thereof, is of great interest as QITE has many practically relevant applications.
Suppose that the initial state has a non-zero overlap with the ground state of $H$, then all components that do not correspond to the ground state are damped exponentially in time during imaginary time evolution. This form of time evolution is, thus, a particularly useful tool to find the ground state of $H$ \cite{VarSITEMcArdle19}.
Furthermore, imaginary time evolution can be used to solve partial differential equations \cite{gonzalezconde2021pricing, fontanela2021quantumpdes, Kubo_2021StochastidDifferential}, to prepare quantum Gibbs states \cite{VarQBMZoufal20, Simon18TheoryVarQSim, MottaQITE20, Temme2011QuantumMS, YungQuantumMetropolis12, WiebeVariationalGibbs2020}, or to solve combinatorial optimization problems \cite{gacon2021simultaneous}.

In order to implement QTE on an actual quantum computer, the time evolution must be translated into a hardware native process. Thus, quantum simulation on a gate-based quantum computer requires a translation into quantum gates which may, e.g., be approximated with Trotterization \cite{Lloyd1073UniversalQuantumSim96, MottaQITE20}. While this approach has many advantages, the resulting quantum circuits can easily become too deep for reliable execution on near-term devices.
\textit{Variational quantum time evolution} (VarQTE)
\cite{VarSITEMcArdle19, Simon18TheoryVarQSim, EfficientVarQuantumSimErrorMinBenjamin17, Endo_2020VarQTEGeneralProcesses} offers an interesting alternative that can simulate quantum time dynamics with (shallow) parameterized quantum circuits. Next to its compatibility with shallow, variational quantum circuits, the method's ability to offload parts of the algorithmic calculations to classical computers makes it a promising candidate for solving interesting QTE problems with near-term devices.
However, VarQTE does rely on a variational approximation and, hence, generally comes with an approximation error. The efficient quantification of this error is crucial to allow for interpretation of the results, and to possibly adapt the simulations.

Error bounds for algorithmic and implementation induced errors of variational quantum real time evolution (VarQRTE) given by the time dependent variational principle~\cite{dirac_1930TDVP} with respect to the trace distance $D_{\mathrm{Tr}}$ are introduced in \cite{EfficientVarQuantumSimErrorMinBenjamin17}. The respective bound depends on the operator norm of $H^2$ which usually scales unfavorably. This issue is resolved in a consecutive work presented in~\cite{Endo_2020VarQTEGeneralProcesses} where efficient error bounds for the variational simulations of general processes--including VarQRTE and variational quantum imaginary time evolution (VarQITE)--are derived for the trace distance.
Furthermore,~\cite{MartinazzoErrorVarQuantumDyn20} introduces an error bound $\epsilon^{\text{PD}}$ for the $\ell_2$-norm between target state and state prepared with VarQRTE based on McLachlan's variational principle \cite{McLachlan64} that suffers from global phase dependence.
Interestingly, the algorithmic error bound for VarQRTE presented in~\cite{Endo_2020VarQTEGeneralProcesses} is equivalent to the bound from~\cite{MartinazzoErrorVarQuantumDyn20}, i.e., $\epsilon^{\text{PD}}$. 

\begin{figure}[h!]
\captionsetup{singlelinecheck = false, format= hang, justification=centerlast, font=footnotesize, labelsep=space}
\begin{center}
\begin{tikzpicture}[scale=1.1]
\node[] at (0,0) {\large{$D_{\mathrm{Tr}}$}};
\node at (0.5,0) {\large{$\leq$}};
\node at (0.9,0) {\large{$B$}};
\node at (1.3,0) {\large{$\leq$}};
\node at (1.7,0) {\large{$\ell_2$}};
\node[rotate=-90] at (0.9,-0.4) {\large{$\leq$}};
\node[rotate=-90] at (1.7,-0.4) {\large{$\leq$}};
\node at (0.9,-0.85) {\large{$\epsilon$}};
\node at (1.75,-0.8) {\large{$\epsilon^{\text{PD}}$}};
\node at (1.3,-0.88) {\large{$\leq$}};
\end{tikzpicture}
\end{center}
\caption{This figure illustrates the inequality relations between the distance metrics and algorithmic error bounds that have been considered in related work respectively are being considered in this work.}
\label{fig:inequalities}
\end{figure}

In this work, we introduce a posteriori error bounds $\epsilon$ for VarQRTE and VarQITE based on the Bures distance $B$ which is agnostic to physically irrelevant global phase mismatches -- a feature that is aligned with the theory of state of the art implementations \cite{Simon18TheoryVarQSim}.
These error bounds are an important progress towards practical quantum simulation verification for VarQTE.
More specifically, the obtained error bounds enable us to efficiently quantify the algorithmic approximation error with respect to the optimal QTE solution.
The additionally required resources to evaluate the bounds are minimal, as most quantities are already known from the variational principle itself.
Furthermore, the bounds are practically easy to implement through numerical integration of an ordinary differential equation (ODE) that is defined by residual quantities stemming from the underlying variational equations. 
Moreover, the new error bounds $\epsilon$ define lower bounds on the phase dependent $\epsilon^{\text{PD}}$ from \cite{Endo_2020VarQTEGeneralProcesses, MartinazzoErrorVarQuantumDyn20} and on the fidelity between prepared and target states due to a direct relation of the Bures metric to the fidelity.
It should also be noted that the Bures metric upper bounds the trace distance.
The Bures metric and $\ell_2$-norm as well as $\epsilon$ and $\epsilon^{\text{PD}}$ are trivially equivalent if the time evolution does not introduce a global phase change or the variational ansatz manages to perfectly capture the global phase change.
The equivalence between trace distance and Bures metric on the other hand only holds if the underlying states are equal up to global phase, i.e., if they are zero. Otherwise, the inequality becomes strict.
It directly follows that we can also derive a bound on $D_{\mathrm{Tr}}$ using $\epsilon$ that is strictly better than $\epsilon^{\text{PD}}$.
The inequality relations of the various metrics and respective error bounds are illustrated in Fig.~\ref{fig:inequalities} and discussed in more detail in Appendix \ref{app:inequalities}.

Our main contributions of this work are summarized as follows: Firstly, this work presents improved algorithmic error bounds $\epsilon$ for VarQTE implementations that are based on McLachlan's variational principle that lower bound existing bounds $\epsilon^\text{PD}$ \cite{MartinazzoErrorVarQuantumDyn20, Endo_2020VarQTEGeneralProcesses}.
We discuss practically relevant aspects considering the integration of the ODE underlying VarQTE. 
Lastly, the practical behavior of the novel error bounds is demonstrated on various numerical examples. We investigate their performance and illustrate the application to concrete settings.

The structure of this work is as follows. First, we explain the concepts of (variational) quantum time evolution in Sec.~\ref{sec:VarQTE}. Then, Sec.~\ref{sec:error_bounds} introduces the a posteriori error bounds for VarQRTE and VarQITE. Furthermore, methods used for the numerical experiments are described in Sec.~\ref{sec:methods} and the respective results are presented in Sec.~\ref{sec:results}. Finally, conclusions and outlook are given in Sec.~\ref{sec:conclusions}.

\section{Variational Quantum Time Evolution}
\label{sec:VarQTE}

VarQTE maps the time evolution of a state $\ket{\psi^*_t} $ onto a variational ansatz state $\ket{\psi^{\omega}_t}$ whose time dependence is projected onto the parameters $\bm{\omega_t}$. To simplify the notation, the time parameter $t$ is dropped from $\bm{\omega}=(\omega_0, \ldots, \omega_k)\in\mathbb{R}^{k+1}$ in the remainder of this work when referring to the ansatz parameters.
More specifically, we consider the current state of the art formulation for pure states based on McLachlan's variational principle \cite{McLachlan64} with a global phase agnostic evolution \cite{Simon18TheoryVarQSim}. 
This formulation directly compensates for terms that arise if the global phase of the state changes during the time evolution.

The state evolution described by the variational principle corresponds to an initial value problem where the underlying \textit{ordinary differential equation} (ODE)~\cite{Tahir-Kheli2018ODEs} is derived from McLachlan's variational principle \cite{McLachlan64}.
We simulate the time evolution by numerically solving the ODE for a set of initial parameter values. It should be noted that the respective formulation not only enables QTE simulations for Hamiltonians given as weighted sum of Pauli operators but also for Hamiltonians which are incompatible with Trotterization such as those given in first quantization \cite{QAlgGridVarQRTE}. 
In the following, real and imaginary time evolution, as well as, the variational implementations are introduced. Notably, we set w.l.o.g. $\hbar=1$. 

\subsection{Variational Quantum Real Time Evolution}
\label{sec:varqrte}
The time-dependent Schr\"odinger equation describes the change of a state $\ket{\psi^*_t}$ under real time evolution 
\begin{align}
    i\ket{\dot\psi^*_t}= H\ket{\psi^*_t},
\end{align}
where we use the time derivative notation
\begin{align}
    \dot a = \frac{\partial a}{\partial t}.
\end{align}
The resulting time-dependent state reads
\begin{align}
\label{eq:VarQRTE_state}
    \ket{\psi^*_t} = \ee^{-iHt}\ket{\psi^*_0}.
\end{align}

The variational approximation of $\ket{\psi^*_t}$ with $\ket{\psi^{\omega}_t}$ is based on the ODE defined by
 \begin{es}
 \label{eq:McLachlanVarQRTE}
      \sum\limits_{j=0}^{k}\mathcal{F}^Q_{ij} \dot\omega_j = \text{Im}\left( C_i -  \frac{\partial \bra{\psi^{\omega}_t}}{\partial \omega_i}\ket{\psi^{\omega}_t}E_t^{\omega}\right),
\end{es}
where $C_i = \frac{\partial \bra{\psi^{\omega}_t}}{\partial \omega_i} H\ket{\psi^{\omega}_t}$, $E_t^{\omega}=\bra{\psi^{\omega}_t}H\ket{\psi^{\omega}_t}$ and
$\mathcal{F}_{ij}^Q$ denotes the $(i,j)$-entry of the Fubini-Study metric \cite{QFIMBraunstein94, meyer2021fisher} given by
\begin{align}
\mathcal{F}^Q_{ij} =
\text{Re}\left(\frac{\partial \bra{\psi^{\omega}_t}}{\partial \omega_i}\frac{\partial \ket{\psi^{\omega}_t}}{\partial \omega_j} - \frac{\partial \bra{\psi^{\omega}_t}}{\partial \omega_i}\proj{\psi^{\omega}_t}\frac{\partial \ket{\psi^{\omega}_t}}{\partial \omega_j}\right). \nonumber
\end{align}
Practically, Eq.~\eqref{eq:McLachlanVarQRTE} tells us that we need to find the parameter update $\boldsymbol{\dot{\omega}}$ which minimizes
\begin{align}
\label{eq:standardODE_varqrte}
    f_{\text{res}}\left(\bm{\omega}\right) = \underset{\boldsymbol{\dot{\omega}}\in\mathbb{R}^{k+1}}{\text{argmin}} \, \left\|\mathcal{F}^Q\boldsymbol{\dot{\omega}} -\text{Im}\left( \bm{C} -  \frac{\partial \bra{\psi^{\omega}_t}}{\partial \bm{\omega}}\ket{\psi^{\omega}_t}E_t^{\omega}\right)\right\|,
\end{align}
with $\bm{C} = \left(C_0, \ldots, C_k\right)$.

We would like to point out that for pure states the Fubini-Study metric is proportional to the quantum Fisher Information matrix.
The derivation of Eq.~\eqref{eq:McLachlanVarQRTE} is presented in Appendix \ref{app:global_phase_agnostic_varqrte} and the efficient evaluation of the respective terms is discussed in Appendix \ref{app:implementation}.
Due to the variational approximation, the gradient $\ket{\dot\psi^{\dot\nu}_t}$ will typically not be exact and, hence, $\| \ket{e_t}\| > 0$
for
\begin{es}
\label{eq:gradientErrorVarQRTE}
     \ket{e_t} := \ket{\dot\psi^{\dot\nu}_t}+ iH\ket{\psi^{\omega}_t}
\end{es}
denoting the gradient error or residual of a single VarQRTE step.

We may also consider our problem from a different angle. Instead of considering Eq.~\ref{eq:standardODE_varqrte}, we may also look for the argument $\boldsymbol{\dot{\omega}}$ which minimizes the residual, i.e.,
\begin{es}
\label{eq:argminODE_varqrte}
  f_{\text{err}}\left(\bm{\omega}\right)=\underset{\boldsymbol{\dot{\omega}}\in\mathbb{R}^{k+1}}{\text{argmin}} \, \|\ket{e_{t}}\|_2 ^2,
\end{es}
where
\begin{align}
\label{eq:qrte_error}
    \norm{\ket{e_{t}}}_2  ^2 &= \Var\left(H\right)_{\psi^{\omega}_t} + \left( \braket{\dot\psi^{\omega}_t|\dot\psi^{\omega}_t} - \braket{\dot\psi^{\omega}_t|\psi^{\omega}_t}\braket{\psi^{\omega}_t|\dot\psi^{\omega}_t}\right)  \nonumber \\
    &\hspace{5mm}- 2\textnormal{Im}\left(\bra{\dot\psi^{\omega}_t}H\ket{\psi^{\omega}_t}-E_t^{\omega} \braket{\dot\psi^{\omega}_t|\psi^{\omega}_t} \right),
\end{align}
for $\Var(H)_{\psi^{\omega}_t} = \bra{\psi^{\omega}_t} H^2\ket{\psi^{\omega}_t} - (E_t^{\omega})^2$ and $\textstyle{
 2\text{Re}(\braket{\psi^{\omega}_t|\dot\psi^{\omega}_t}) = \frac{\partial\braket{\psi^{\omega}_t|\psi^{\omega}_t}}{\partial t} = 0}.$

Since the time-dependence of $\ket{\psi^{\omega}_t}$ is encoded in the real parameters $\bm{\omega}$,
Eq.~\eqref{eq:qrte_error} can be further rewritten as
\begin{align}
\label{eq:error_rqte_var}
    \|\ket{e_t}\|_2^2 &= \Var(H)_{\psi^{\omega}_t} + \sum\limits_{i, j}\dot \omega_i \dot \omega_j \mathcal{F}^Q_{ij} \nonumber \\ 
    &\hspace{5mm}- 2\sum_i\dot\omega_i\text{Im}\left(C_i -  \frac{\partial \bra{\psi^{\omega}_t}}{\partial \omega_i}\ket{\psi^{\omega}_t}E_t^{\omega}\right)
\end{align}
using
\begin{align}
     &\text{Im}\left(\bra{\dot \psi^{\omega}_t}H\ket{\psi^{\omega}_t}-E_t^{\omega}\braket{\dot \psi^{\omega}_t|\psi^{\omega}_t}\right) \nonumber \\ 
     &\hspace{10mm}= \sum_i\dot\omega_i\text{Im}\left( C_i -  \frac{\partial \bra{\psi^{\omega}_t}}{\partial \omega_i}\ket{\psi^{\omega}_t}E_t^{\omega}\right) 
\end{align}
and
\begin{align} \label{eq_Christa_reform1}
   \braket{\dot\psi^{\omega}_t|\dot\psi^{\omega}_t} - \braket{\dot\psi^{\omega}_t|\psi^{\omega}_t}\braket{\psi^{\omega}_t|\dot\psi^{\omega}_t} = \sum_{i,j}\dot \omega_i \dot \omega_j \mathcal{F}^Q_{ij}.
\end{align}
All terms required to compute $\|\ket{e_t}\|_2$ can be evaluated with the techniques presented in Appendix \ref{app:implementation}.

We would like to point out that the solutions to Eq.~\eqref{eq:standardODE_varqrte} and Eq.~\eqref{eq:argminODE_varqrte} are analytically equivalent but the numerical behavior may differ. More specifically, the simulation results presented in Sec.~\ref{sec:results} present a more stable behavior of the latter.

\subsection{Variational Quantum Imaginary Time Evolution}
\label{sec:varqite}
 Imaginary time evolution of a quantum state is mathematically described by the normalized, Wick-rotated Schr\"odinger equation
\begin{align}
\label{eq:wick_schroedinger}
    \ket{\dot\psi^*_t} = \left( E^*_t\mathds{1} - H\right)\ket{\psi^*_t},
\end{align}
where $E^*_t = \bra{\psi^*_t} H\ket{\psi^*_t}$ corresponds to the system energy.
In the remainder of this work, the notation for $E^*_t\mathds{1} - H$ is simplified to $E^*_t - H$.
The state evolution reads
\begin{es}
\label{eq:VarQITE_state}
    \ket{\psi^*_t} = \frac{\ee^{-Ht}\ket{\psi^*_0}}{\sqrt{\bra{\psi^*_0}\ee^{-2Ht}\ket{\psi^*_0}}}\, .
    \end{es}

Analogously to Sec.~\ref{sec:varqrte}, we simulate the existence of an explicit global phase $\ee^{-i\nu}$ \cite{Simon18TheoryVarQSim, VarSITEMcArdle19} and, thereby, avoid the addition of a physically irrelevant phase gate.

Solving 
\begin{es}
\label{eq:McLachlanVarQITE}
\sum\limits_{j=0}^k \mathcal{F}^Q_{ij} \dot\omega_j= - \text{Re}\left(C_i\right),
\end{es}
for $\boldsymbol{\dot\omega}$ leads to an ODE
which describes the evolution of the ansatz parameters in terms of the parameter updates that minimize 
\begin{align}
\label{eq:standardODE_varqite}
        f_{\text{res}}\left(\bm{\omega}\right) = \underset{\boldsymbol{\dot{\omega}}\in\mathbb{R}^{k+1}}{\text{argmin}} \, \left\|\mathcal{F}^Q\boldsymbol{\dot{\omega}}+\text{Re}\left( C_i\right)\right\|.
\end{align}
The derivation of Eq.~\eqref{eq:McLachlanVarQITE} is presented in Appendix \ref{app:global_phase_agnostic_varqite} and details on the evaluation of the individual terms are given in Appendix \ref{app:implementation}.

As before, the state gradients are likely to be inexact due to the variational approximation such that $\|\ket{e_t}\|>0$ for the gradient error
\begin{es}
\label{eq:grad_error_varqite}
    \ket{e_t}:=\ket{\dot\psi^{\dot\nu}_t}  - \Big(E_t^{\omega}-H\Big)\ket{\psi^{\omega}_t}.
\end{es}

Eq.~\eqref{eq:grad_error_varqite} again motivates an alternative VarQITE ODE formulation that aims at finding the parameter updates $\boldsymbol{\dot{\omega}}$ as
\begin{align}
\label{eq:argminODE_varqite}
  f_{\text{err}}\left(\bm{\omega}\right)=\underset{\boldsymbol{\dot{\omega}}\in\mathbb{R}^{k+1}}{\text{argmin}} \, \|\ket{e_{t}}\|_2 ^2,
\end{align}
for
\begin{align}
\label{eq:et_imag}
    \|\ket{e_{t}}\|_2  ^2=   \Var(H)_{\psi^{\omega}_t} 
    &+ \braket{\dot\psi^{\omega}_t|\dot\psi^{\omega}_t} - \braket{\dot\psi^{\omega}_t|\psi^{\omega}_t}\braket{\psi^{\omega}_t|\dot\psi^{\omega}_t} \nonumber  \\
    & +2\mathrm{Re}\big(\!\bra{\dot\psi^{\omega}_t}H\ket{\psi^{\omega}_t}\!\big),
\end{align}
using that
\begin{align}
    2\text{Re}\left(\braket{\psi^{\omega}_t|\dot\psi^{\omega}_t}\right) = \frac{\partial\braket{\psi^{\omega}_t|\psi^{\omega}_t}}{\partial t} = 0.
\end{align}

Since the time-dependence of $\ket{\psi^{\omega}_t}$ is encoded in the parameters $\bm{\omega}$, we can rewrite Eq.~\eqref{eq:et_imag} 
\begin{align}
\label{eq:et_imag_eval}
\|\ket{e_{t}}\|_2  ^2 
= \Var(H)_{\psi^{\omega}_t} \!+\! \sum_{i,j}\dot \omega_i \dot \omega_j \mathcal{F}^Q_{ij} \!+\! 2\sum_i\dot\omega_i\text{Re}(C_i), \nonumber
\end{align}
where we employ the fact that
\begin{es}
\text{Re}\big(\!\bra{\dot\psi^{\omega}_t}H\ket{\psi^{\omega}_t}\!\big) = \sum_i\dot\omega_i\text{Re}(C_i),
\end{es}
as well as Eq.~\eqref{eq_Christa_reform1}.
The efficient evaluation of $\|\!\ket{e_t}\!\|_2^2 $ employs the techniques presented in Appendix \ref{app:implementation}.

\section{Error Bounds}
\label{sec:error_bounds}
In this section, we prove global phase agnostic error bounds for VarQTE.
Let $\ket{\psi^{\omega}_t}$ be the state prepared by the variational algorithm at time $t$ and denote the ideal target state by $\ket{\psi^*_t}$. 
To formalize an error bound, we want to use a global phase agnostic metric which describes the distance between two quantum states.
A popular distance measure is the fidelity \cite{nielsen10} given by $|\braket{\psi^{\omega}_t|\psi^*_t}|^2$. Unlike the $\ell_2$-norm, the fidelity is invariant to changes in the global phases.  Since the global phase is physically irrelevant, 
this is a desired property for a meaningful quantum state distance measure. Although the fidelity itself does not correspond to a metric, it may be used to define the Bures metric~\cite{HayashiQuantumInfo06}, i.e.,
\begin{align}
	&B\left(\ket{\psi^{\omega}_t}, \ket{\psi^*_t} \right)= \nonumber \\ 
	&\hspace{20mm} \sqrt{\braket{\psi^{\omega}_t|\psi^{\omega}_t}+\braket{\psi^*_t|\psi^*_t}-2|\braket{\psi^{\omega}_t|\psi^*_t}|},
 \end{align}
 where the states $\ket{\psi^{\omega}_t}$ and $\ket{\psi^*_t}$ are not necessarily normalized. It should be noted that we use the simplified notation $B\left(\ket{\psi^{\omega}_t}, \ket{\psi^*_t} \right):=B\left(\proj{\psi^{\omega}_t}, \proj{\psi^*_t} \right)$.

Our goal is, now, to prove a bound of the form
\begin{align}
	B\left(\ket{\psi^{\omega}_t}, \ket{\psi^*_t} \right) \leq \epsilon_t,
\end{align}
for an error term $\epsilon_t$ that can be evaluated efficiently in practice.
Interestingly, the error bounds with respect to the Bures metric are a direct consequence of the phase agnostic ODE formulation of VarQTE presented in the previous section.
Dropping the VarQTE terms that compensate for potential global phase changes leads to error bounds for the $\ell_2$-norm $\epsilon^{\text{PD}}$.
The aforementioned relation of the Bures metric to the fidelity
\begin{align}
    |\braket{\psi^{\omega}_t|\psi^*_t}| \geq 1 - \frac{\epsilon_t^2}{2},
\end{align}
 implies that the relevant range of $\epsilon_t$ is $\epsilon_t\in\left[0, \sqrt{2}\right]$ for normalized $\ket{\psi^{\omega}_t}$ and $\ket{\psi^*_t}$.
If the error bound estimate lies outside of this interval, then the fidelity and error can be clipped to $0$ and  $\sqrt{2}$, respectively.

\subsection{Variational Quantum Real Time Evolution}
\label{subsec:error_qrte}
In~\cite{MartinazzoErrorVarQuantumDyn20} the authors derive an error bound for a VarQRTE formulation that does not include the global phase compensating terms. Due to the global phase dependent nature of the VarQRTE ODE, the resulting error bound presents an upper bound to the $\ell_2$-error.
We, now, align the theory with the global phase independent formulation of VarQRTE and derive a corresponding error bound for the Bures metric which helps to avoid that a physically irrelevant mismatch in the global phase influences the bound. The proof is given in Appendix~\ref{app:proofThm1} and a comparison between $\ell_2$-norm and Bures metric error bound is given in Appendix~\ref{app:comparison}.

\begin{theorem} \label{thm_VQRT}
For $T>0$ and $\epsilon_0=0$, let $\ket{\psi^*_T}$ be the exact solution to Eq.~\eqref{eq:VarQITE_state} and $\ket{\psi^{\omega}_T}$ correspond to the VarQRTE approximation. Then
\begin{align}
 B\left(\ket{\psi^*_T}, \ket{\psi^{\omega}_T}\right) \leq \epsilon_{T} := \int_{0}^{T} \norm{\ket{e_t}}_2 \di t
\end{align}
for 
\begin{align}
        \|\ket{e_t}\|^2_2 &= \Var(H)_{\psi^{\omega}_t} + \sum\limits_{i, j}\dot \omega_i \dot \omega_j \mathcal{F}^Q_{ij} \nonumber \\ 
    &\hspace{5mm}- 2\sum_i\dot\omega_i \I \left(C_i -  \frac{\partial \bra{\psi^{\omega}_t}}{\partial \omega_i}\ket{\psi^{\omega}_t}E_t^{\omega}\right) . \label{eq:varqrte_error_grad}
\end{align}
\end{theorem}
It should be noted that the error bound is compatible with practical VarQRTE implementations which use, e.g., regularized least squares methods or pseudo-inversion methods to solve the system of linear equations from Eq.~\eqref{eq:McLachlanVarQRTE}.


\subsection{Variational Quantum Imaginary Time Evolution}
\label{subsec:error_qite}
This section introduces an upper-bound to the Bures metric between the target state $\ket{\psi^*_t}$ given by Eq.~\eqref{eq:wick_schroedinger} and $\ket{\psi^{\omega}_t}$ prepared with VarQITE.
The proof is given in Appendix~\ref{app:proofThm2}.
\begin{theorem} \label{thm_VQITE}
For $T>0$ and $\epsilon_0=0$, let $\ket{\psi^*_T}$ be the exact solution to Eq.~\eqref{eq:VarQITE_state} and $\ket{\psi^{\omega}_T}$ be the simulation implemented using VarQITE.  Then
\begin{align}
 B\left(\ket{\psi^*_T}, \ket{\psi^{\omega}_T}\right) \leq \epsilon_T :=\int_{0}^{T} \norm{\ket{e_t}}_2 \di t,
\end{align}
for 
\begin{align}
\|\ket{e_{t}}\|^2_2 
&= \Var(H)_{\psi^{\omega}_t} \!+\! \sum_{i,j}\dot \omega_i \dot \omega_j \mathcal{F}^Q_{ij} \nonumber \\ 
    &\hspace{5mm}\!+\! 2\sum_i\dot\omega_i \R (C_i). 
 \label{eq:varqite_error_grad}
\end{align}
\end{theorem}
This error bound is also independent of a potential physically irrelevant global phase mismatch between prepared and target state. Furthermore, the bound is compatible with  implementations which use numerical techniques to solve the SLE given in Eq.~\eqref{eq:McLachlanVarQITE}.

\section{Implementation}
\label{sec:methods}
To ensure a stable VarQTE implementation, it is vital to choose the correct settings. The possible choices with their advantages and disadvantages, respectively, are explained next.

\subsection{ODE Solvers}
\label{sec:odesolvers}
The ODE underlying VarQTE is solved using numerical integration. This can lead to an additional error term. 

Let $\ket{\psi_t^*}$ denote the target state, $\ket{\psi^{\omega}_t}$ the prepared state, and $\ket{\psi_t'}$ the state that we would prepare if we could take infinitesimally small time steps and, thus, integrate the ODE exactly.
Then, the error bounds derived in Sec.~\ref{sec:error_bounds} capture the errors induced by the variational method, i.e.,
\begin{es}
	B\left(\ket{\psi_t^*}, \ket{\psi_t'}\right) \leq \epsilon_t.
\end{es}
The triangle inequality gives
\begin{align}
	B\left(\ket{\psi^*_t}, \ket{\psi^{\omega}_t}\right) &\leq B\left(\ket{\psi^*_t}, \ket{\psi'_t}\right)+B\left(\ket{\psi'_t}, \ket{\psi^{\omega}_t}\right).
\end{align}
The term $B(\ket{\psi'_t}, \ket{\psi^{\omega}_t})$ is generally unknown and the error bounds from Sec.~\ref{sec:error_bounds} only hold if $B(\ket{\psi'_t}, \ket{\psi^{\omega}_t}) \ll 1$ such that 
\begin{es}
	B\left(\ket{\psi^*_t}, \ket{\psi^{\omega}_t}\right)  
	\!\approx\! B\left(\ket{\psi^*_t}, \ket{\psi'_t}\right) 
	\!\leq\! \epsilon_t.
\end{es}

ODE solvers, such as the \emph{forward Euler} method, which operate with a fixed step size may induce large errors in the numerical simulations if the time steps are not chosen sufficiently small.
The forward Euler method evaluates the gradient $\boldsymbol{\dot\omega_t}$ and propagates the underlying variable for $n_T$ time steps according to a predefined step size, i.e.,
\begin{es}
   \bm{\omega_{T}} = \bm{\omega_0} + \sum\limits_{k=0}^{n_T}  \delta_t \boldsymbol{\dot{\omega}_{t_k}}
\end{es}
with $t_{n_T}=T$ and the step-size $\delta_t=t_{k+1}-t_k$.
In contrast, \emph{Runge-Kutta} methods evaluate additional supporting points and compute a parameter update using an average of these points, thereby, truncating the local update error. 
Combining two Runge-Kutta methods of different order but using the same supporting points allows to define efficient adaptive step-size ODE solvers which ensure that the local step-by-step error is small and, thus, that the dominant part of the error is coming from the variational approximation.
The results in Sec.~\ref{sec:results} illustrate this aspect on the example of the forward Euler method \cite{Griffiths2010NumericalODE} with fixed step  size and an explicit Runge-Kutta method of order 5(4) (RK54) method from SciPy \cite{2020SciPy-NMeth} that uses additional interpolation points as well as an adaptive step size to minimize the step-by-step integration errors.
We refer the interested reader to an introductory book on numerical ODE solvers such as \cite{Griffiths2010NumericalODE}.

\subsection{ODE Definition}
The SLE underlying McLachlan's variational principle, given in Eq.~\eqref{eq:McLachlanVarQRTE} and Eq.~\eqref{eq:McLachlanVarQITE}, are prone to being ill-conditioned and may, thus, only be solvable approximately with a numerical technique such as regularized least squares or pseudo-inversion. The commonly used regularization schemes, as well as, the pseudo-inversion can be seen as small perturbations which are not necessarily in accordance with the physics of the system. This in turn can lead to inappropriate parameter updates. In the following, we shall refer to the ODE definition based on $ f_{\text{res}}$ as \textit{residual ODE}.
The alternative ODE definition $f_{\text{err}}$ --  which shall be referred to as \textit{gradient error ODE} -- is analytically equivalent to solving $ f_{\text{res}}$. However, the simulation results in Sec.~\ref{sec:results} show that the numerical behavior differs.
In fact, the experiments reveal that the gradient error ODE can lead to significantly better numerical stability. 

The simulations employ the SciPy COBYLA optimizer \cite{2020SciPy-NMeth} to find $\boldsymbol{\dot\omega}$ in $f_{\text{err}}$ where the initial point is chosen
as the numerical solution to the SLE given in Eq.~\eqref{eq:McLachlanVarQRTE} and Eq.~\eqref{eq:McLachlanVarQITE}, respectively.

\subsection{Error Bound Evaluation}
\label{sec:ode_form}
To enable a reliable error bound evaluation, we jointly evolve the state parameters and the error bounds. 
More explicitly, we extend the parameter  ODE to
\begin{align}
        \begin{pmatrix}
       \boldsymbol{\dot \omega}_t \\
    \dot \epsilon_t
    \end{pmatrix}= 
        \tilde f\left(\bm{\omega_t}, \epsilon_t\right),
\end{align}
with  $\bm{\omega}_0$ being set, $\epsilon_0 = 0$ by assumption, and 
\begin{align}
    \tilde f\left(\bm{\omega_t}, \epsilon_t\right)=\begin{pmatrix}
        f\left(\bm{\omega_t}\right) \\
        \norm{\ket{e_t}}_2
    \end{pmatrix},
\end{align}
with $\norm{\ket{e_t}}_2$ from Eqs.~\eqref{eq:varqrte_error_grad} and \eqref{eq:varqite_error_grad} for the real and imaginary case, respectively. Furthermore, $ f\left(\bm{\omega_t}\right)$ is either chosen as $ f_{\text{res}}\left(\bm{\omega_t}\right)$ or $f_{\text{err}}\left(\bm{\omega_t}\right)$.
This formulation has the advantage that the error bound directly reflects the propagation of the evolution and that adaptive step size ODE solvers also consider the changes in the error bounds.

\section{Simulation Results}
\label{sec:results}

\begin{figure}[h!]
\captionsetup{singlelinecheck = false, format= hang, justification=centerlast, font=footnotesize, labelsep=space}
\begin{center}
\begin{tikzpicture}
\node at (0,0){\includegraphics[width=0.8\linewidth]{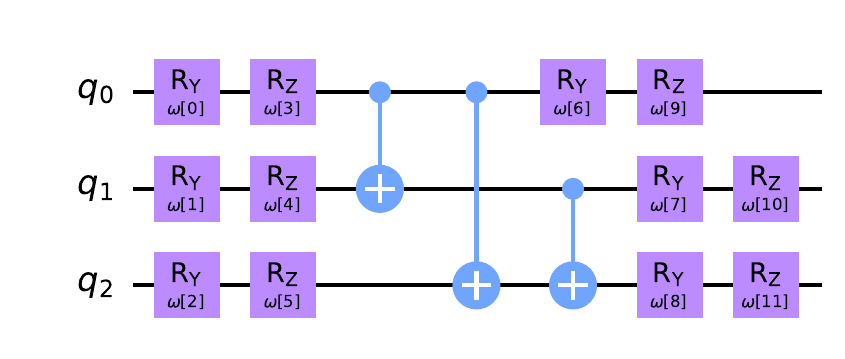}};
\node at (-4.4, 0) {$\ket{\psi_0}$};
\draw[decorate, thick, decoration = {brace, amplitude=10pt}] (-3.5,-1.) --  (-3.5,1.);
\end{tikzpicture}
\end{center}
\caption{This quantum circuit corresponds to the \emph{EfficientSU2} ansatz in  Qiskit's \cite{qiskit_2019} circuit library and is chosen as ansatz for the experiments presented in this work. It consists of layers of $RY$ and $RZ$ rotations and a $CX$ entanglement block which is chosen according to the \textit{full} layout. The number of repetitions is set to $1$.}
\label{fig:effsu2}
\end{figure}

In this, we demonstrate the efficiency of the error bounds derived in Sec.~\ref{sec:error_bounds}, as well as, the impact of the implementation details discussed in Sec.~\ref{sec:methods} and present an example that illustrates the importance of phase agnostic VarQTE error bounds.
Unless otherwise stated, the experiments prepare $\ket{\psi^{\omega}_t}$ with an ansatz as shown in Fig.~\ref{fig:effsu2}, adjusted to the number of qubits $n$ given by the respective Hamiltonian:
\begin{enumerate}[(i)]

\item An illustrative example is considered with
\begin{es}
    H_{\text{illustrative}} = Z\otimes X + X\otimes Z + 3 Z\otimes Z.
\end{es}
Hereby, the evolution time is $T=1$ and the initial parameters are chosen such that all parameters are set to $0$ except for the parameters of the last layer of $RY$ rotations which are chosen to be $\pi/2$. This gives $\ket{\psi_0}=\ket{++}$.

\item The well-studied Ising model with a transverse magnetic field on an open chain is investigated, see, e.g., \cite{Calabrese_2012TransveresIsing}, i.e.,
\begin{es}
\label{eq:ising}
     H_{\text{Ising}} = -J\left(\sum\limits_{i,j}Z_i\otimes  Z_j + g\sum\limits_j X_j\right),
\end{es}
where $J=-\frac{1}{2}$ and $g=-\frac{1}{2}$.
While the following section includes examples with $3$ qubits, additional results with $10$ qubits are presented in Appendix \ref{app:10qubits}.
The evolution time is again set to $T=1$ and the initial parameters are all $0$ except for the parameters of the last layer of $RZ$ gates which are chosen at random in $(0, \frac{\pi}{2}]$ such that $\ket{\psi_0}=\ee^{-i\gamma}\ket{000}$ with $\gamma\in\mathbb{R}$.
Notably, we avoid the initial state $\ket{\psi_0}=\ket{000}$ to circumvent getting stuck in a local minima.

\item The two qubit hydrogen molecule approximation given in \cite{VarSITEMcArdle19} is studied, with
    \begin{align}
        H_{\text{hydrogen}} 
        &= 0.2252\,I\otimes I + 0.5716\,Z\otimes Z \nonumber \\
        &\hspace{3mm}+ 0.3435\,I\otimes Z  - 0.4347\,Z\otimes I \nonumber \\
        &\hspace{3mm}+ 0.0910\,Y\otimes Y + 0.0910\,X\otimes X.     \label{eq:hydrogen}
    \end{align}
    Again, the evolution time is set to $T=1$ and the initial parameters are chosen such that the initial state is $\ket{\psi_0}=\ket{++}$, i.e., all parameters are $0$ except for the last layer of $RY$ rotations which are given as $\pi/2$.
\end{enumerate}

\subsection{Variational Quantum Real Time Evolution}
\label{sec:varqrte_results}

\begin{figure}[!htb]
    \centering
    \captionsetup{singlelinecheck = false, format= hang,justification=centerlast, font=footnotesize, labelsep=space}
    \begin{tikzpicture}
\node at (-1, 0.5) {\textbf{VarQRTE: $H_{\text{illustrative}}$ with different ODE solvers}};
\node[inner sep=0pt, anchor=north west] at (-5.5, -0.12) {\includegraphics[width=0.25\textwidth]{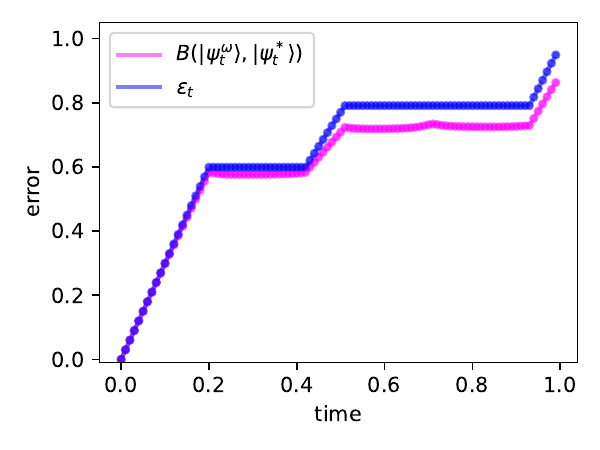}};
\node[anchor=north west] at (-1,0) {    \includegraphics[width=0.25\textwidth]{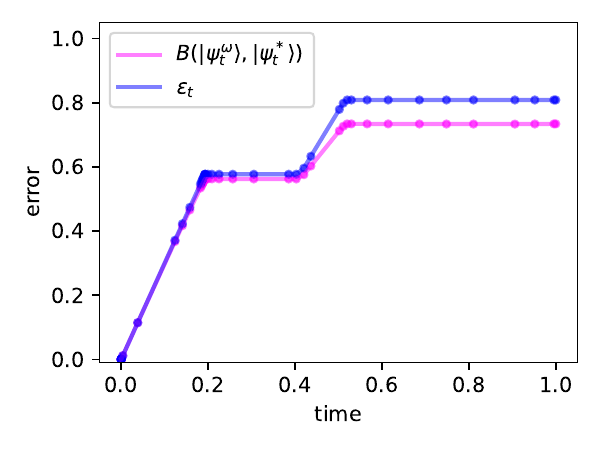}};
\node[anchor=north west] at (-5.5, -3.7) {\includegraphics[width=0.25\textwidth]{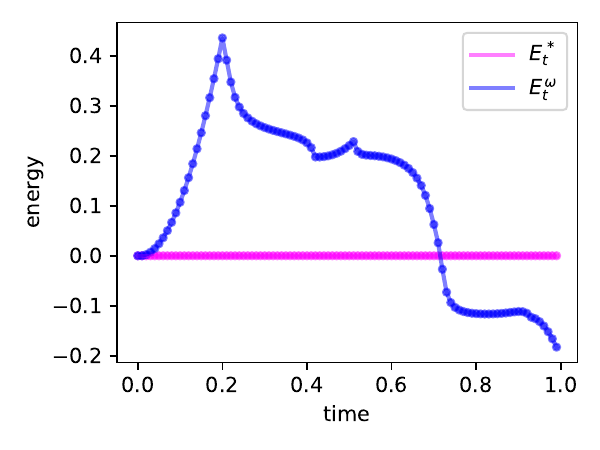}};
\node[anchor=north west] at (-1,-3.7) {\includegraphics[width=0.25\textwidth]{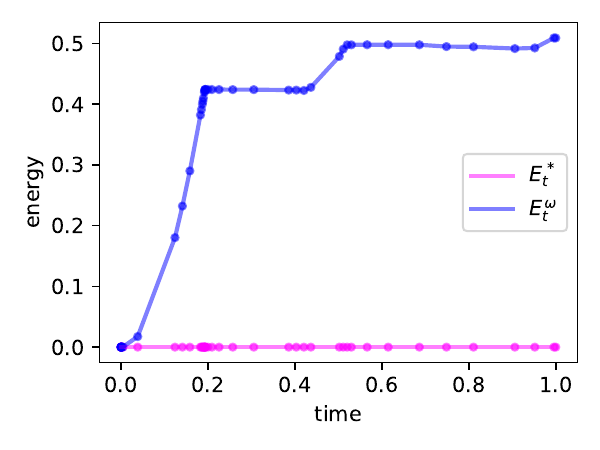}};
\node at (-3.2, 0) {(a) State Error (Euler, $ f_{\text{res}}$) };
\node at (1.3, 0) {(b)  State Error (RK54, $ f_{\text{res}}$) };
\node at (-3.2, -3.7) {(c)  Energy (Euler, $ f_{\text{res}}$) };
\node at (1.3, -3.7) {(d) Energy (RK54, $ f_{\text{res}}$) };
\end{tikzpicture}
    \caption{VarQRTE for $\ket{\psi_0}=\ket{++}$, $H_{\text{illustrative}}$ and $T=1$  with the residual ODE. (a), (c) employ forward Euler. (b), (d) use RK54. -  (a), (b) illustrate the error bounds $\epsilon_t$ and the actual Bures metric between $\ket{\psi^*_t}$ and  $\ket{\psi^{\omega}_t}$.  (c), (d) show the corresponding energies.}
    \label{fig:illustrative_varqrte}
\end{figure}
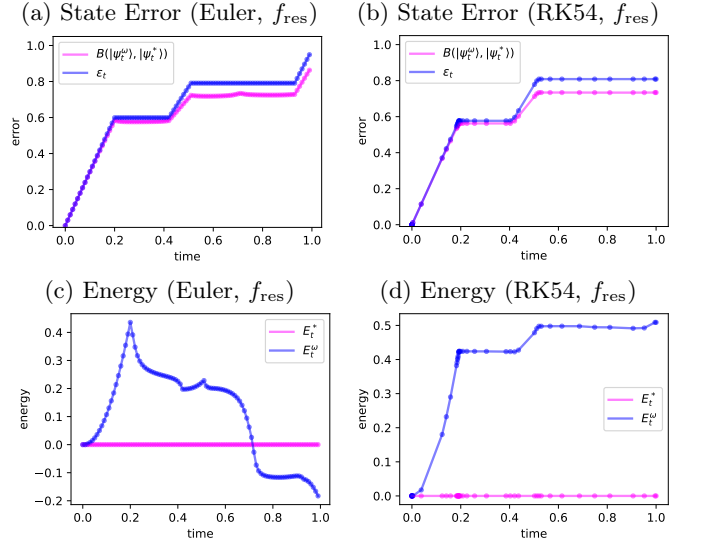
\begin{figure}[!ht]
    \centering
    \captionsetup{singlelinecheck = false, format= hang, justification=centerlast, font=footnotesize, labelsep=space}
    \begin{tikzpicture}
\node at (-1, 0.5) {\textbf{VarQRTE: $H_{\text{Ising}}$ with different ODE definitions}};
\node[inner sep=0pt, anchor=north west] at (-5.5, -0.12) {\includegraphics[width=0.25\textwidth]{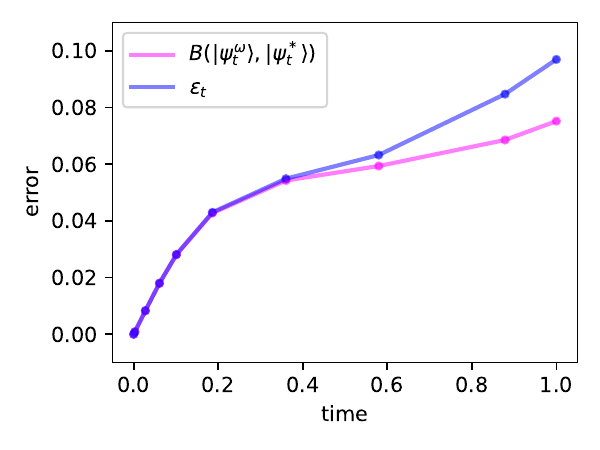}};
\node[anchor=north west] at (-1,0) {    \includegraphics[width=0.25\textwidth]{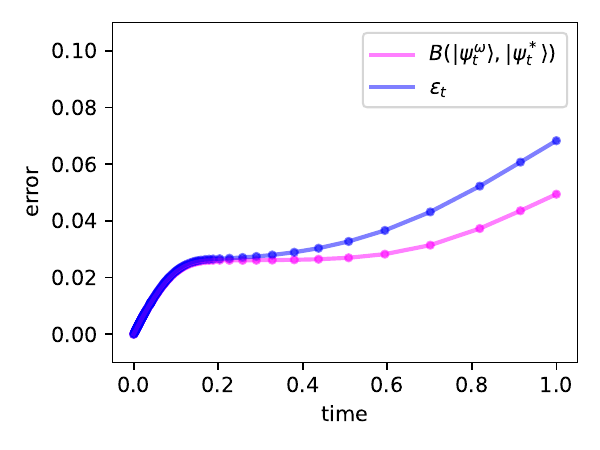}};
\node[anchor=north west] at (-5.5, -3.7) {\includegraphics[width=0.25\textwidth]{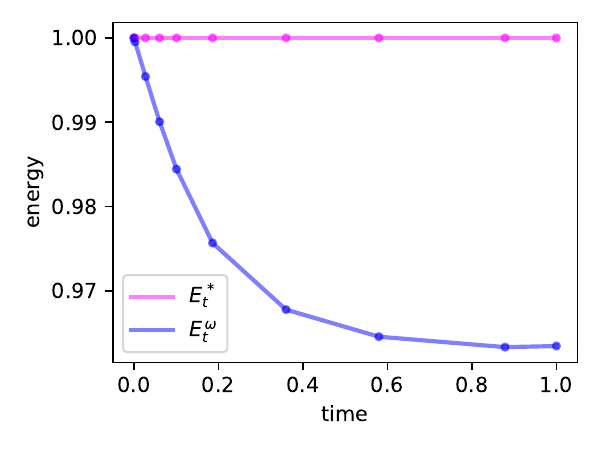}};
\node[anchor=north west] at (-1,-3.7) {\includegraphics[width=0.25\textwidth]{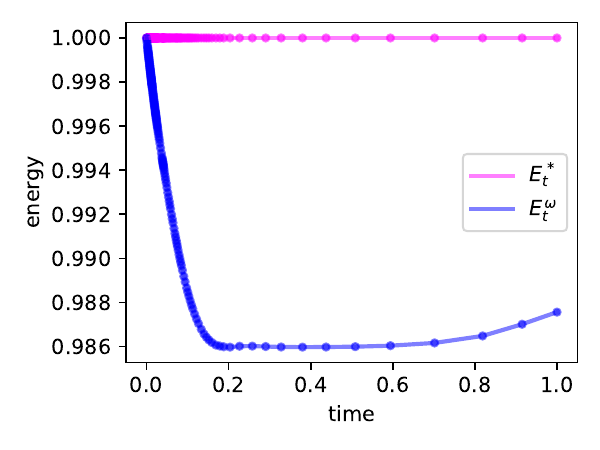}};
\node at (-3.2, 0) {(a) State Error (RK54, $ f_{\text{res}}$)};
\node at (1.3, 0) {(b) State Error (RK54, $f_{\text{err}}$)};
\node at (-3.2, -3.7) {(c) Energy (RK54, $ f_{\text{res}}$)};
\node at (1.3, -3.7) {(d) Energy (RK54, $f_{\text{err}}$)};
\end{tikzpicture}
    \caption{VarQRTE for $\ket{\psi_0}=\ee^{-i\alpha}\ket{000}$, $H_{\text{Ising}}$ and $T=1$ with RK54. (a), (c) are based on the residual ODE.  (b), (d) use the gradient error ODE. - (a), (b) illustrate the error bounds $\epsilon_t$ and the actual Bures metric. (c), (d) show the energies corresponding to prepared and target state.}
    \label{fig:ising_varqrte}
\end{figure}
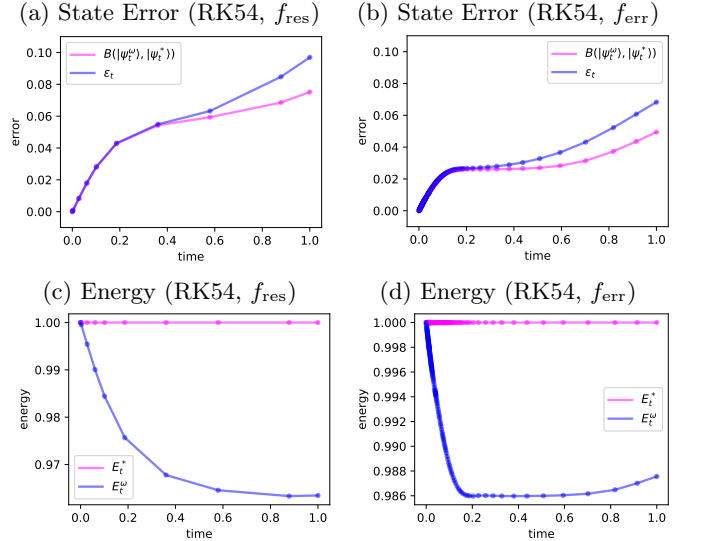

In the following, we present a set of numerical experiments and investigate the error bounds for VarQRTE with a particular focus on the comparison of different ODE formulations and solvers.

Firstly, we apply the forward Euler method with $100$ time steps as well as an adaptive step size RK54 ODE solver to the illustrative example using the residual ODE. The parameter propagation given by $f_{\text{res}}$ is solved using a least square solver provided by NumPy \cite{Numpy2020} with a cut-off ratio for small singular values of $0.001$.
The results shown in Fig.~\ref{fig:illustrative_varqrte} illustrate that the error bounds are very tight and, thus, relevant for practical accuracy estimations.  
Furthermore, one can see that RK54 achieves a state preparation with less error compared to Forward Euler--which is reflected in the error bounds--as well as smaller fluctuations in the system energy while using significantly less time steps. The plateaus are due to exact local gradients, i.e., $\|\ket{e_{t}}\|_2 = 0$.
Furthermore, we would like to point out that the energy should actually be preserved for a real time evolution under a time-independent Hamiltonian but McLachlan's variational principle does not guarantee energy preservation.

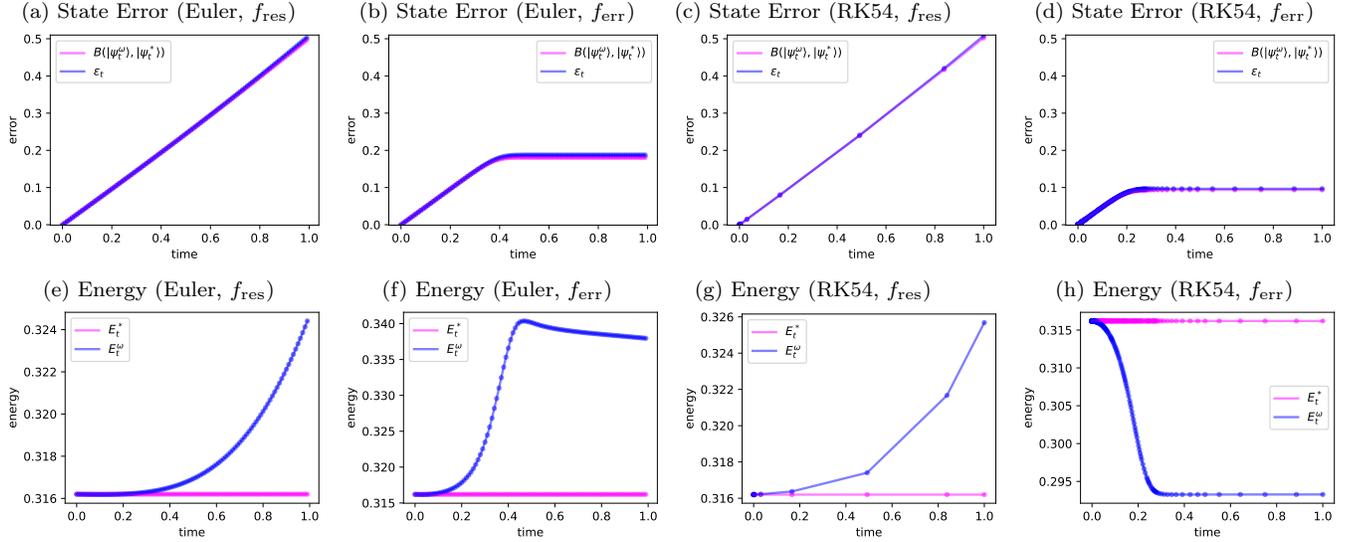
\begin{figure*}[!ht]
    \captionsetup{singlelinecheck = false, format= hang, justification=centerlast, font=footnotesize, labelsep=space}
    \centering
    \begin{tikzpicture}
     \node at (3.5, 0.5) {\textbf{VarQRTE: $H_{\text{hydrogen}}$ with different ODE solvers and different ODE types}};
     \node[anchor=north west] at (-5.5, 0) {\includegraphics[width=0.25\textwidth]{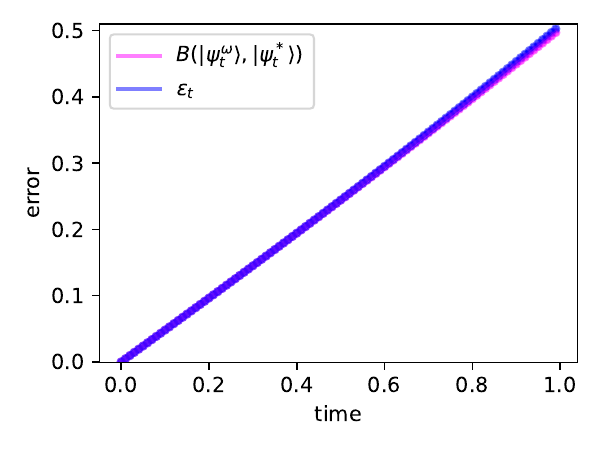}};
\node[anchor=north west] at (-1, 0) {\includegraphics[width=0.25\textwidth]{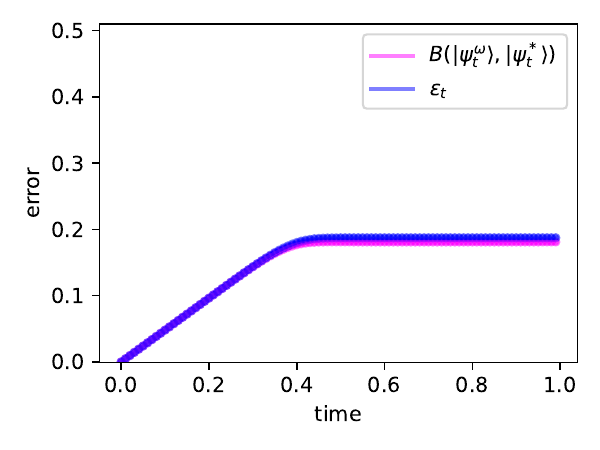}};
\node[anchor=north west] at (3.5,0) {\includegraphics[width=0.25\textwidth]{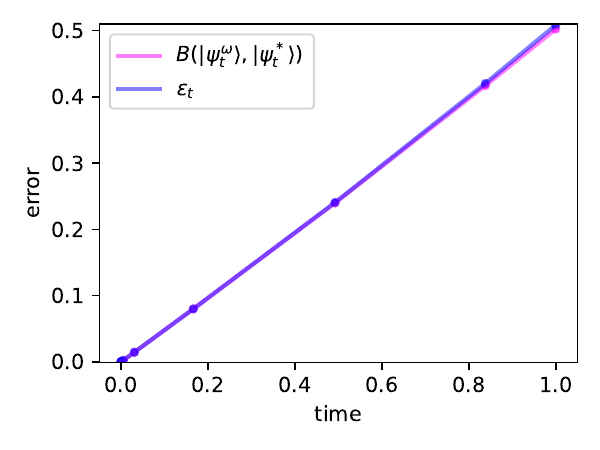}};
\node[anchor=north west] at (8, 0) {\includegraphics[width=0.25\textwidth]{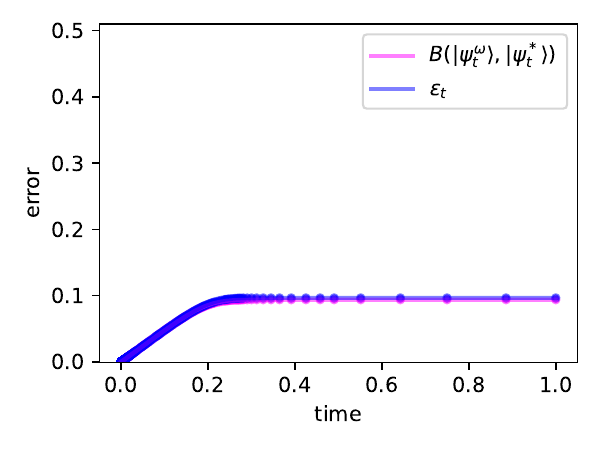}};
\node[anchor=north west] at (-5.5, -3.7) {\includegraphics[width=0.25\textwidth]{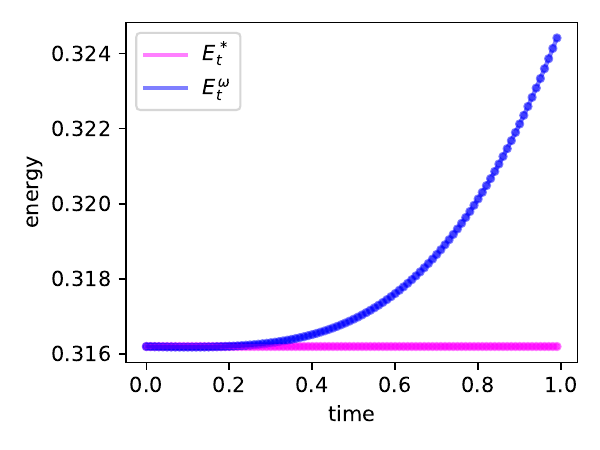}};
\node[anchor=north west] at (-1, -3.7) {\includegraphics[width=0.25\textwidth]{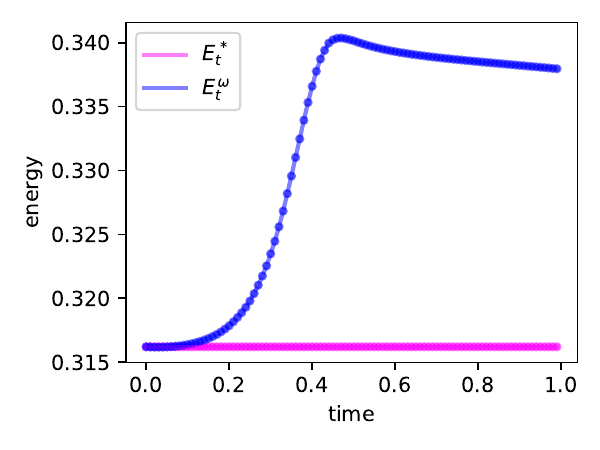}};
\node[anchor=north west] at (3.5,-3.7) {\includegraphics[width=0.25\textwidth]{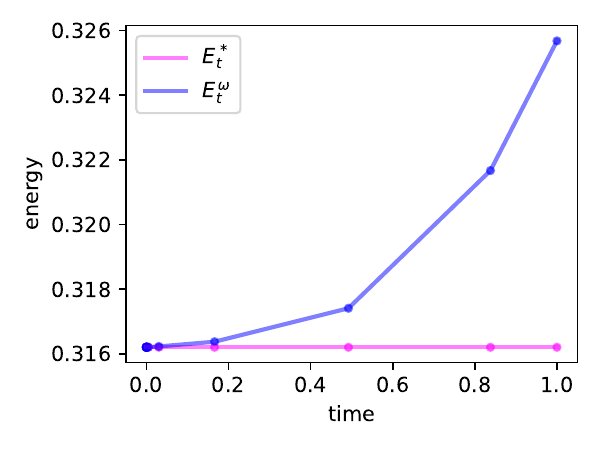}};
\node[anchor=north west] at (8, -3.7) {\includegraphics[width=0.25\textwidth]{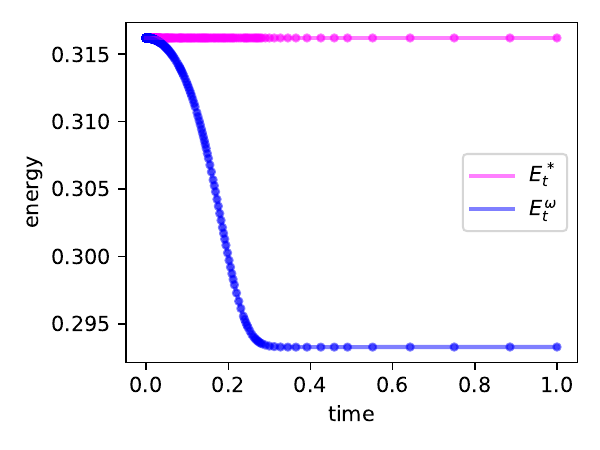}};
\footnotesize{
\node at (-3.2, 0) {(a) State Error (Euler, $ f_{\text{res}}$) };
\node at (1.3, 0) {(b) State Error (Euler, $f_{\text{err}}$)};
\node at (5.5, 0) {(c) State Error (RK54, $ f_{\text{res}}$)};
\node at (10.3, 0) {(d) State Error (RK54, $f_{\text{err}}$)};
\node at (-3.2, -3.7) {(e) Energy (Euler, $ f_{\text{res}}$)      };
\node at (1.3, -3.7) {(f) Energy (Euler, $f_{\text{err}}$)      };
\node at (5.5, -3.7) {(g) Energy (RK54, $ f_{\text{res}}$)      };
\node at (10.3, -3.7) {(h) Energy (RK54, $f_{\text{err}}$)       };}
\end{tikzpicture}
    \caption{VarQRTE for $\ket{\psi_0}=\ket{++}$, $H_{\text{hydrogen}}$ and $T=1$. (a), (b), (e), (f) employ a forward Euler solver. (c), (d), (g), (h) use an RK54 ODE solver. (a), (c), (e), (g) use the residual ODE. (b), (d), (f), (h) rely on the gradient error ODE. -
    (a), (b), (c), (d) illustrate the error bounds $\epsilon_{t}$ and the true Bures metric. (e), (f), (g), (h) show the respective energies $E_t^{\omega}$ and $E_t^*$.}
    \label{fig:hydrogen_varqrte_energy}
\end{figure*}

Next, we compare the impact of $f_{\text{res}}$ compared to $f_{\text{err}}$ on the example of an Ising model using RK54.
Here, $f_{\text{res}}$ and $f_{\text{err}}$ are solved with a least square solver provided by NumPy \cite{Numpy2020} and an additional a regularization on the Fubini-Study metric. More explicitly, we use $\mathcal{\tilde F} = \mathcal{F} + \gamma\mathds{1}$ for a small $\gamma$.
The initial points for the optimization of the gradient error ODE are chosen as the solution to the respective SLE at time $t$.
Fig.~\ref{fig:ising_varqrte} presents the Bures metrics, as well as, the respective bounds for the prepared $\ket{\psi^{\omega}_t}$ and  the target state $\ket{\psi^*_t}$. The errors show that the gradient error ODE leads to smaller  errors than the residual ODE. Furthermore, it can be seen that also the system energy changes less for the former.
Furthermore, Appendix \ref{app:10qubits} shows error bound results for an Ising model with $10$ qubits. These experiments highlight the potential of the error bounds to be applicable for systems with larger dimensions.


Lastly, the error bounds for the hydrogen Hamiltonian from Eq.~\eqref{eq:hydrogen} are compared the residual and gradient error ODE as well as forward Euler and RK54 ODE solvers. 
In this case, $f_{\text{res}}$ and $f_{\text{err}}$ are solved using ridge regression, also known as Tikhonov regularization, from SciKit \cite{scikit-learn2011}. 
This method is also used to compute the initial values for the gradient error ODE formulation.
The results are presented in Fig.~\ref{fig:hydrogen_varqrte_energy}. Notably, the experiment which uses RK54 and the gradient error ODE leads to the best results, i.e., the smallest state error as well as error bound. 
In general, one can see that the gradient error ODE achieves better errors compared to the residual ODE, the error seems to saturate for the former while it keeps increasing with the latter.
Furthermore, RK54 improves the errors, as well as, the error bounds for both ODE definitions while using significantly less time steps.
We would like to point out that the setting which gives to the smallest error $\epsilon_t$ does not necessarily lead to the smallest discrepancy between $E_t^{\omega}$ and $E_t^*$, as can be seen when comparing the RK54 results.

To sum this up, the numerical results reveal that the error bounds represent good estimates for the actual errors. The experiments indicate further that an adaptive step size ODE solver such as RK54 significantly improves the simulation results while reducing the computational costs. Moreover, it was shown that replacing the residual ODE by the gradient error ODE has also a a positive influence on the simulation accuracy. 
Lastly, the results reveal that the lack of energy conservation in McLachlan's variational principle can lead to significant energy fluctuations.

\subsection{Variational Quantum Imaginary Time Evolution}
\begin{figure}[!ht]
    \centering
    \captionsetup{singlelinecheck = false, format= hang,justification=centerlast, font=footnotesize, labelsep=space}
    \begin{tikzpicture}
\node at (-1, 0.5) {\textbf{VarQITE: $H_{\text{illustrative}}$ with different ODE solvers}};
\node[inner sep=0pt, anchor=north west] at (-5.5, -0.12) {\includegraphics[width=0.25\textwidth]{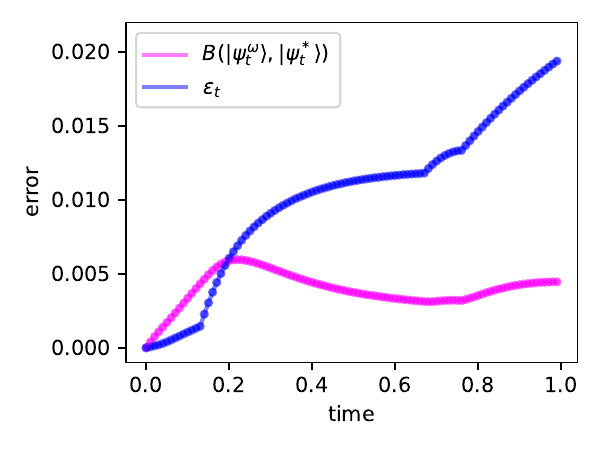}};
\node[anchor=north west] at (-1,0) {    \includegraphics[width=0.25\textwidth]{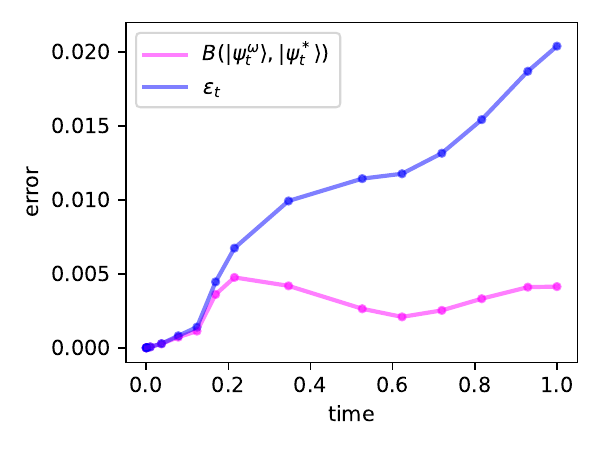}};
\node[anchor=north west] at (-5.5, -3.7) {\includegraphics[width=0.25\textwidth]{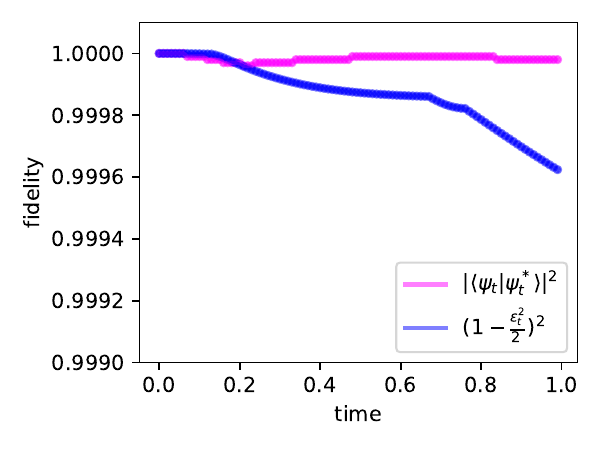}};
\node[anchor=north west] at (-1,-3.7) {\includegraphics[width=0.25\textwidth]{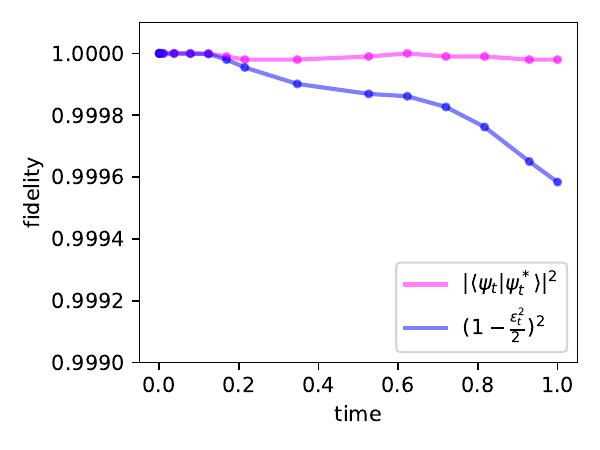}};
\node at (-3.2, 0) {(a) State Error (Euler, $ f_{\text{res}}$) };
\node at (1.3, 0) {(b)  State Error (RK54, $ f_{\text{res}}$) };
\node at (-3.2, -3.7) {(c)  Fidelity (Euler, $ f_{\text{res}}$) };
\node at (1.3, -3.7) {(d) Fidelity (RK54, $ f_{\text{res}}$) };
\end{tikzpicture}
    \caption{VarQITE for $\ket{\psi_0}=\ket{++}$, $H_{\text{illustrative}}$ and $T=1$  with the residual ODE  (a), (c) are computed using Forward Euler. (b), (d) employ  RK54. - (a), (b) illustrate the error bound $\epsilon_{t}$ and the true Bures metric. Furthermore, (c), (d) show the true fidelities and fidelity bounds.}
\label{fig:illustrative_qite}
\end{figure}
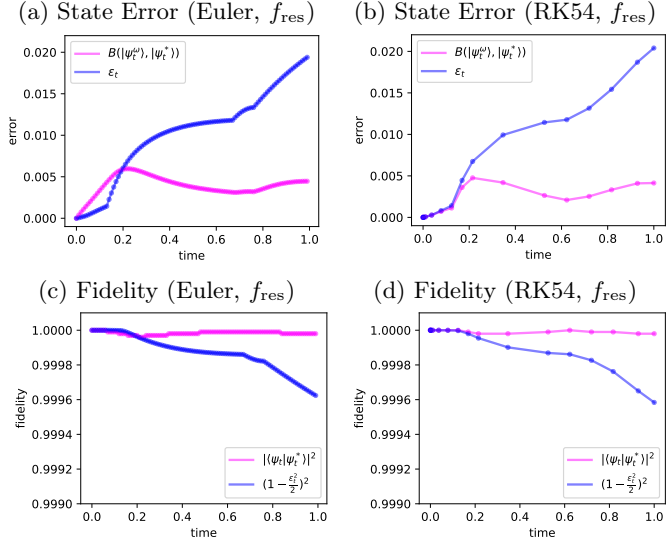
Next, we investigate the practical behavior of the error bounds for VarQITE.
First, the outcomes using the residual ODE with forward Euler as well as RK54 are compared for $H_{\text{illustrative}}$. For all of the following experiments, $f_{\text{res}}$ and $f_{\text{err}}$ are evaluated with ridge regression.
The results shown in Fig.~\ref{fig:illustrative_qite} provide an example of the potentially insufficient numerical integration accuracy of forward Euler. More explicitly, the integration error outweighs the algorithmic error and, hence, the error bounds are at first lower than the actual error. The application of RK54 in comparison reduces the error in the integration. The fidelity plots translate the error bound into a physically easy to interpret distance metric.

Next, the performance of the residual and gradient error ODE formulation is compared on the example of $H_{\text{Ising}}$ using RK54 for an evolution over time $T=5$ using a least square solver provided by NumPy \cite{Numpy2020} with a cut-off ratio for small singular values of $0.001$. Fig.~\ref{fig:qite_ising} illustrates the sensitivity of the variational errors to the underlying ODE formulation. More explicitly, the errors and corresponding bounds are more than twice as big for the VarQITE implementation based on $f_{\text{res}}$ compared to the one based on $f_{\text{err}}$. The larger state error also manifests itself in a deviation of the system energy.
Appendix \ref{app:10qubits} shows progress towards a study of error bounds for larger system dimensions with a $10-$qubit Ising model.

In all previous examples the gradient error ODE formulation lead to better performance. Next, we are going to investigate an example for $H_{\text{hydrogen}}$ where the residual ODE turns out to be the preferable method.
We tested a variety of settings and found that while the residual ODE was leading to reasonable results in most cases, the gradient ODE often lead to spiky gradient errors and, eventually, to large state errors.
One set of results is visualized in Fig.~\ref{fig:hydrogen}. It is clearly illustrated that the residual ODE formulation leads to a significantly better evolution with errors and error bounds that are on the order of $10^{-3}$ compared to a maximal error $\approx 0.56$ around $T=5$ and a bound converging to $\sqrt{2}$ for $f_{\text{err}}$.
Interestingly, in the latter case the energy with respect to the prepared state $E_t^{\omega}$ first strongly deviates from $E_t^*$ but finally reaches similar values again. This indicates that VarQITE provides a promising method for ground state search where the evolution does not necessarily need to be followed perfectly at all times.
The presented results were computed using again a regularization on the Fubini-Study metric: $\mathcal{\tilde F} = \mathcal{F} + \gamma\mathds{1}$ for a small $\gamma$.

\begin{figure}[!ht]
    \centering
    \begin{tikzpicture}
 \node at (-1, 0.5) {\textbf{VarQITE: $H_{\text{Ising}}$ with different ODE definitions}};
\node[inner sep=0pt, anchor=north west] at (-5.5, -0.12) {\includegraphics[width=0.25\textwidth]{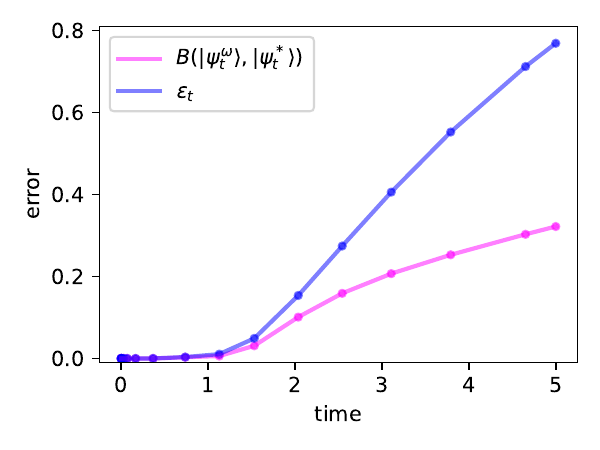}};
\node[anchor=north west] at (-1,0) {    \includegraphics[width=0.25\textwidth]{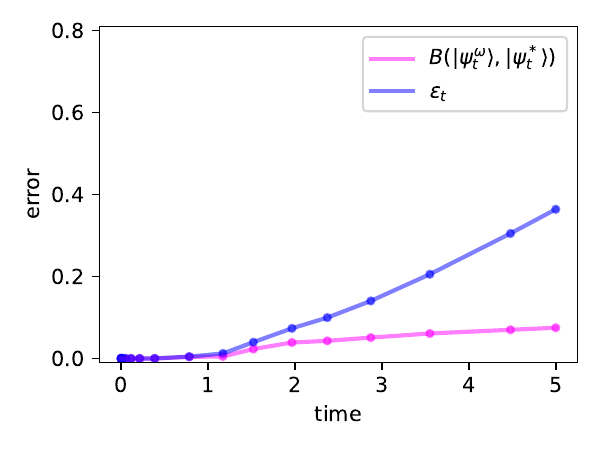}};
\node[anchor=north west] at (-5.5, -3.7) {\includegraphics[width=0.25\textwidth]{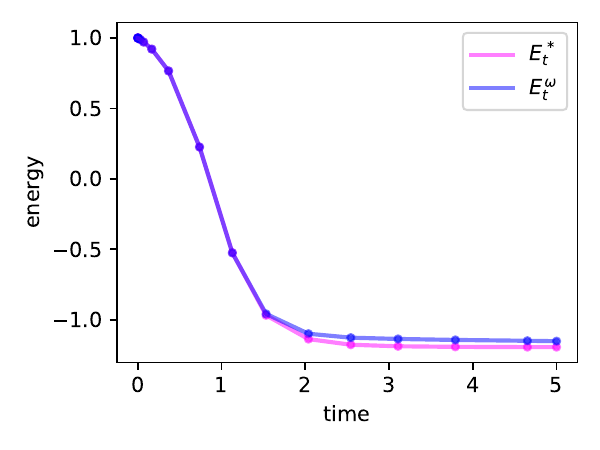}};
\node[anchor=north west] at (-1,-3.7) {\includegraphics[width=0.25\textwidth]{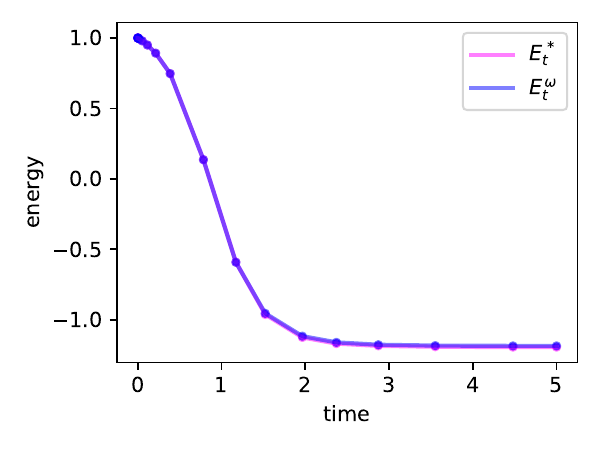}};
\node at (-3.2, 0) {(a) State Error (RK54, $ f_{\text{res}}$) };
\node at (1.3, 0) {(b) State Error (RK54, $f_{\text{err}}$)};
\node at (-3.2, -3.7) {(c) Energy (RK54, $ f_{\text{res}}$)};
\node at (1.3, -3.7) {(d) Energy (RK54, $f_{\text{err}}$)};
\end{tikzpicture}
        \captionsetup{singlelinecheck = false, format= hang, justification=centerlast, font=footnotesize, labelsep=space}
        \caption{VarQITE for $\ket{\psi_0}=\ee^{-i\gamma}\ket{000}$, $H_{\text{Ising}}$ and $T=5$ with RK54. (a), (c) employ the residual ODE. (b), (d) use the gradient error ODE. (a), (b) illustrate the error bound $\epsilon_{t}$, as well as, the actual Bures metric. (c), (d) present the corresponding energy evolution.}
            \label{fig:qite_ising}
\end{figure}
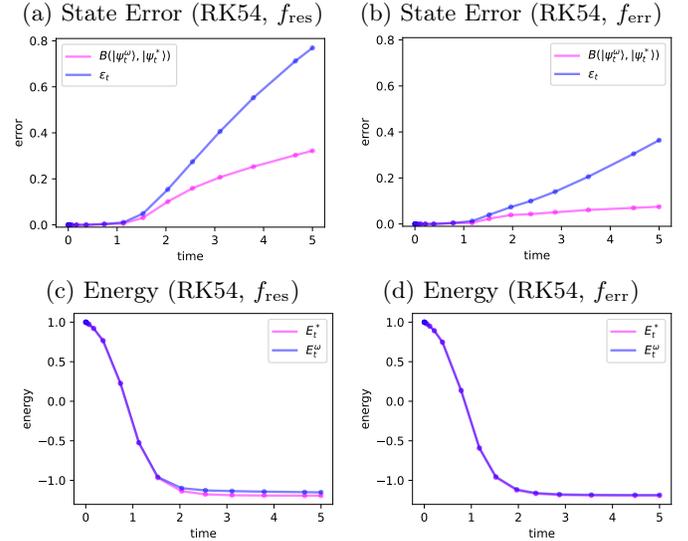
\begin{figure}[!ht]
    \centering
    \begin{tikzpicture}
 \node at (-1, 0.5) {\textbf{VarQITE: $H_{\text{hydrogen}}$}};
\node[anchor=north west] at (-5.5, -0.) {\includegraphics[width=0.25\textwidth]{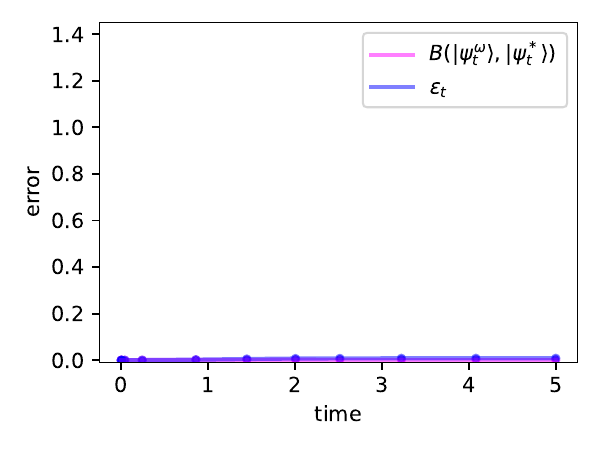}};
\node[anchor=north west] at (-5.5,-3.7) {\includegraphics[width=0.25\textwidth]{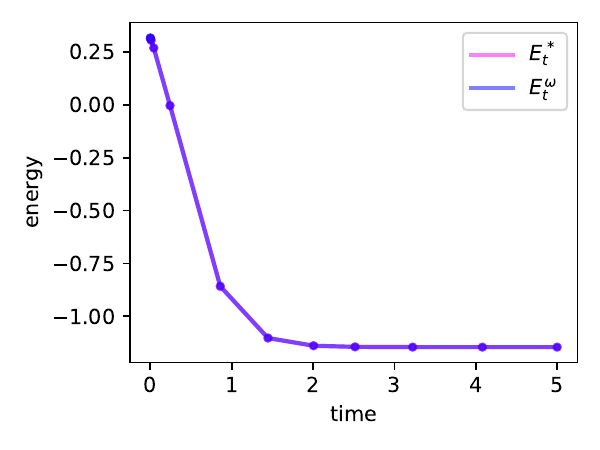}};
\node[anchor=north west] at (-1, 0) {\includegraphics[width=0.25\textwidth]{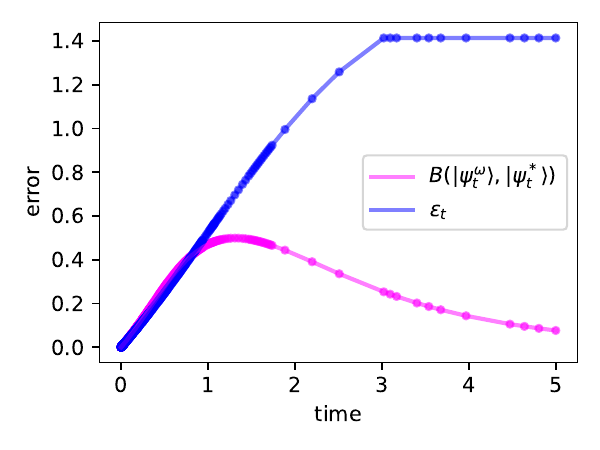}};
\node[anchor=north west] at (-1, -3.7) {\includegraphics[width=0.25\textwidth]{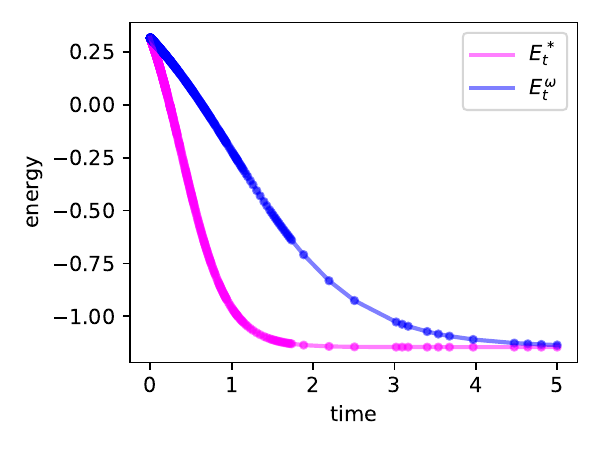}};
\node at (-3.2, 0) {(a) State Error (RK54, $f_{\text{res}}$)};
\node at (1.3, 0) {(b) State Error (RK54, $f_{\text{err}}$)};
\node at (-3.2, -3.7) {(c) Energy (RK54, $f_{\text{res}}$)};
\node at (1.3, -3.7) {(d) Energy (RK54, $f_{\text{err}}$)};
\end{tikzpicture}
        \captionsetup{singlelinecheck = false, format= hang, justification=centerlast, font=footnotesize, labelsep=space}
        \caption{VarQITE for $\ket{\psi_0}=\ket{++}$, $H_{\text{hydrogen}}$ and $T=5$.  All plots are based on RK54 and either the (a), (c) the residual or the (b), (d) gradient error ODE. - (a), (b) illustrate the error bound $\epsilon_{t}$, as well as, the actual Bures metric. Furthermore, (c), (d) present the system energy $E_t^{\omega}$ corresponding to the prepared state and the energy $E_t^*$ corresponding to the target state.}
            \label{fig:hydrogen}
\end{figure}
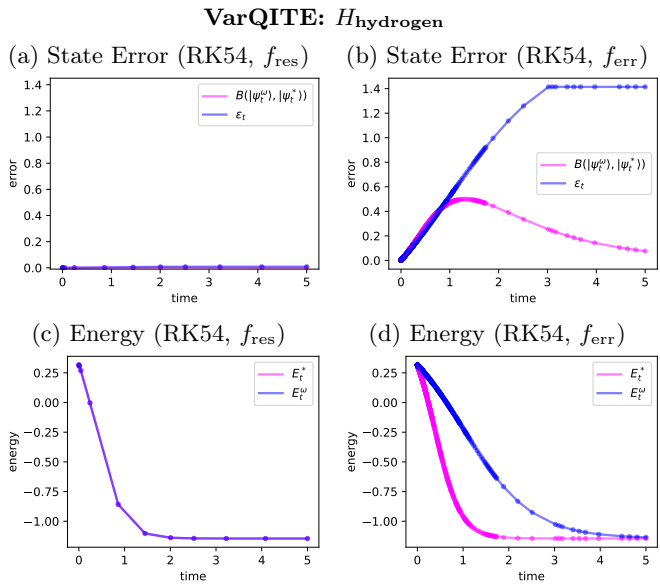

\subsection{Global Phase Dependence}
\label{app:comparison}
\begin{figure}[!hb]
    \centering
    \captionsetup{singlelinecheck = false, format= hang, justification=centerlast, font=footnotesize, labelsep=space}
    \begin{tikzpicture}
\node at (-1, 0.5) {\textbf{VarQRTE: $H=Z$ global phase dependence}};
\node[inner sep=0pt, anchor=north west] at (-5.5, -0.12) {\includegraphics[width=0.25\textwidth]{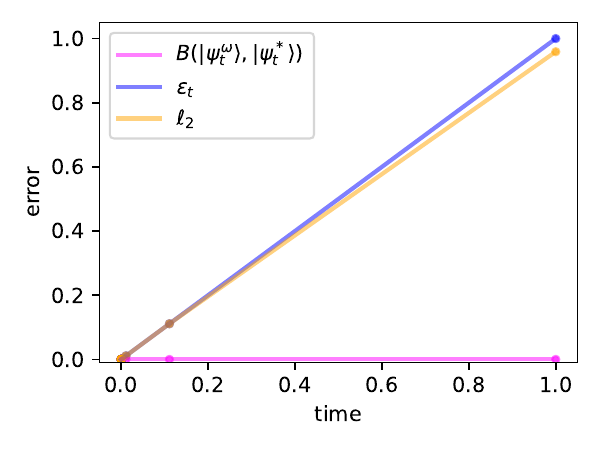}};
\node[anchor=north west] at (-1,0) {    \includegraphics[width=0.25\textwidth]{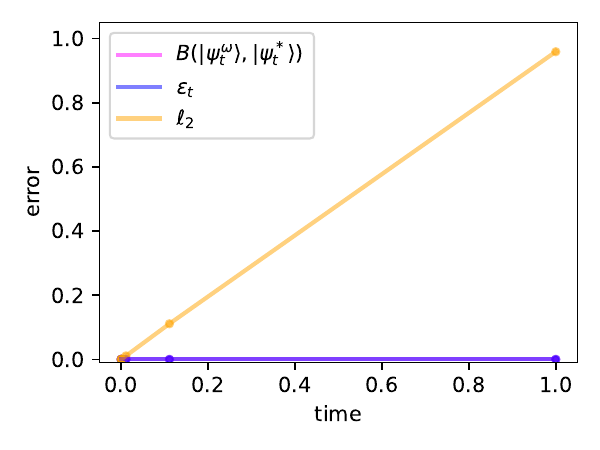}};
\node[anchor=north west] at (-5.5, -3.7) {\includegraphics[width=0.25\textwidth]{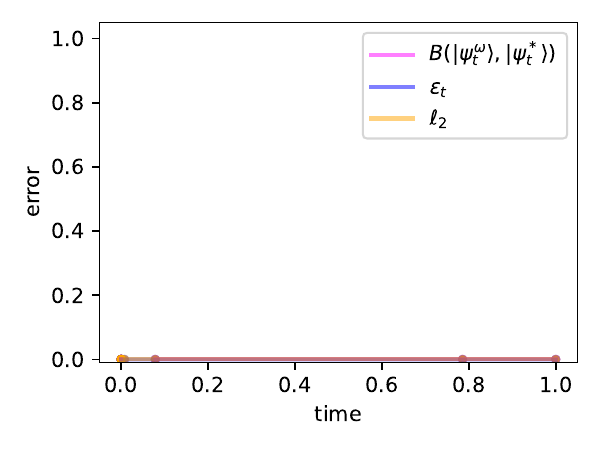}};
\node[anchor=north west] at (-1,-3.7) {\includegraphics[width=0.25\textwidth]{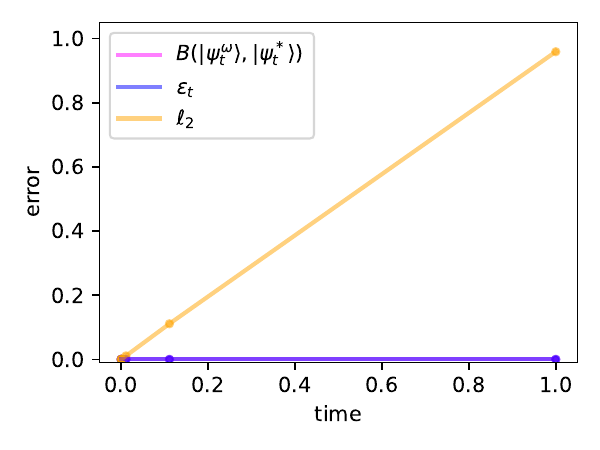}};
\node at (-3.4, 0) {(a) Phase dep. ($RY$)};
\node at (1.3, 0) {(b) Phase agn. ($RY$)};
\node at (-3.2, -3.7) {(c) Phase dep. ($RZRY$)};
\node at (1.3, -3.7) {(d) Phase agn. ($RZRY$)};
\end{tikzpicture}
    \caption{VarQRTE error bounds for $\ket{\psi_0}=\ket{1}$, $H=Z$ and $T=1$ with RK54. (a), (c) are based on the phase dependent definition of the VarQRTE ODE and the error bound  (b), (d) use the phase agnostic definition. - (a), (b) employs ansatz~\ref{ansatz:RY}. (c), (d) employs ansatz~\ref{ansatz:RZRY}.}
    \label{fig:phase_fix_comp}
\end{figure}
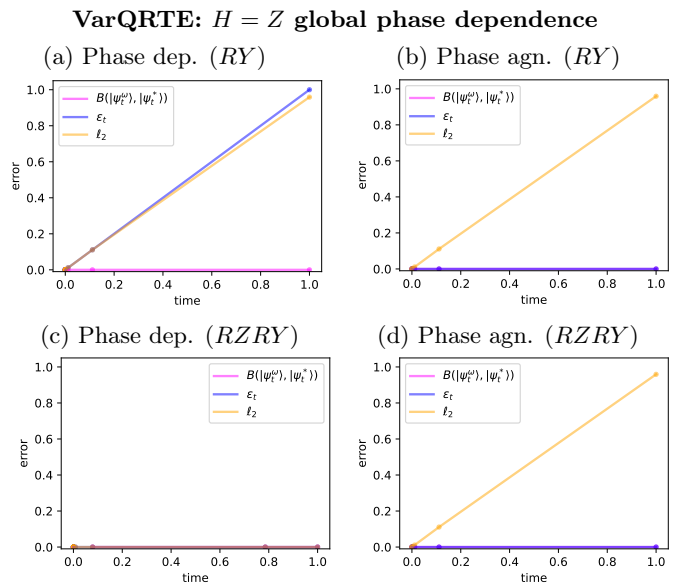

We employ an illustrative example to compare the error bound for VarQRTE derived in this work to the error bound presented in \cite{MartinazzoErrorVarQuantumDyn20} and, thereby, highlight the importance of phase agnostic metrics.
Given the Hamiltonian $H=Z$, we consider the evolution of the initial state $\ket{\psi_0}=\ket{1}$ by $e^{-iHT}$ for $T=1$.
It should be noted that this time evolution solely affects the global phase of $\ket{\psi_0}$ that, as discussed in the main text, is physically irrelevant.
We run VarQRTE with two exemplary ans\"atze
\begin{enumerate}
\item $\ket{\psi_t^{\omega}}=RY\left(\omega_0\right)\ket{1}$ which does not enable the representation of a global phase change, and \label{ansatz:RY}
\item $\ket{\psi_t^{\omega}}=RY\left(\omega_1\right)RZ\left(\omega_0\right)\ket{1}$ which does enable the representation of a global phase change. \label{ansatz:RZRY}
\end{enumerate}
Fig.~\ref{fig:phase_fix_comp} shows the error bounds for VarQRTE using the phase agnostic (phase dependent) McLachlan's variational principle leading to an upper bound $\epsilon_t$ ($\epsilon^{PD}_t$) for the Bures metric ($\ell_2-$norm). The figures also present the exact errors.
All experiments are run with an RK54 ODE solver using the residual ODE with all ansatz parameters being initially set to $0$.

Given the $RY$ ansatz, the $\ell_2-$norm deviates significantly from the Bures metric. While the error bound for the phase dependent formulation $\epsilon^{PD}_t$ leads to large values the phase agnostic bound $\epsilon_t$ directly reflects that the physics of the system do not change.
The results further illustrate that the $RZRY$ ansatz enables the mitigation of the above problem by training an additional parameter to match the global phase change induced by the evolution. 
The phase agnostic formulation can, thus, help to avoid the implementation of an additional gate and parameter -- whose training may potentially induce errors -- while capturing the physics of the problem.

\subsection{Hardware Simulation}
\label{app:HW}

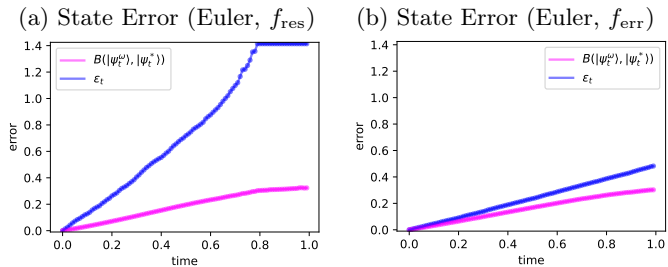
\begin{figure}[!ht]
    \centering
    \captionsetup{singlelinecheck = false, format= hang, justification=centerlast, font=footnotesize, labelsep=space}
    \begin{tikzpicture}
\node at (-1, 0.5) {\textbf{Noisy VarQRTE: $H_{\text{Ising}}$ with different ODE definitions}};
\node[inner sep=0pt, anchor=north west] at (-5.5, -0.12) {\includegraphics[width=0.25\textwidth]{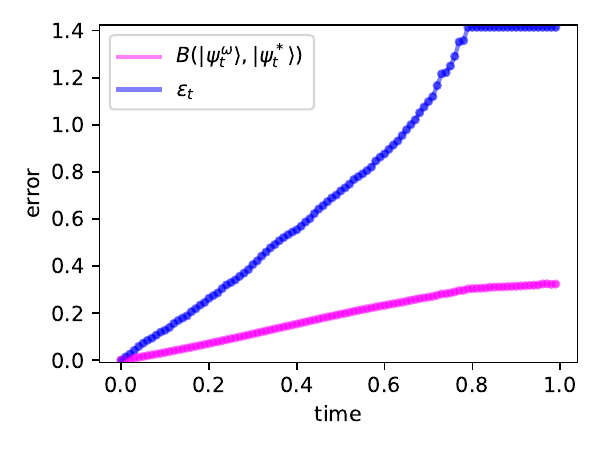}};
\node[anchor=north west] at (-1,0) {    \includegraphics[width=0.25\textwidth]{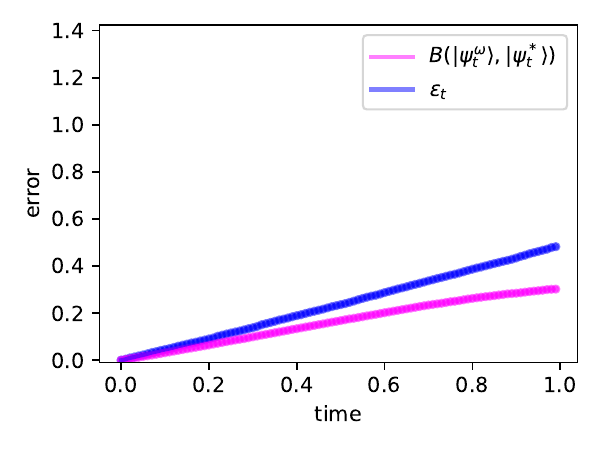}};
\node at (-3.2, 0) {(a) State Error (Euler, $ f_{\text{res}}$)};
\node at (1.3, 0) {(b) State Error (Euler, $f_{\text{err}}$)};
\end{tikzpicture}
    \caption{VarQRTE for $\ket{\psi_0}=\ee^{-i\alpha}\ket{000}$, $H_{\text{Ising}}$ and $T=1$ with Forward Euler. (a) is based on the residual ODE.  (b) uses the gradient error ODE. Both illustrations show the error bounds $\epsilon_t$ and the actual Bures metric.}
    \label{fig:ising_noisy_varqrte}
\end{figure}

The experiments presented so far are run with ideal simulations. However, it is of course also important to understand how the error bounds would perform if the respective quantities would be evaluated with quantum hardware that is affected by physical noise.
As a first study of the robustness of our bounds to hardware noise, we test a VarQRTE experiment for $H_{\text{Ising}}$ described in Sec.~\ref{sec:results} with a noisy simulation for run-time $T=1$. More explicitly, we employ a \textit{density matrix simulator} with a noise model that represents the physical errors of the \textit{IBM Quantum Auckland} backend to evaluate the system variance $\Var(H)_{\psi^{\omega}_t}$, the QFI $\mathcal{F}^Q_{ij}$ and the quantity $\text{Im}\left(C_i -  \frac{\partial \bra{\psi^{\omega}_t}}{\partial \omega_i}\ket{\psi^{\omega}_t}E_t^{\omega}\right)$ given in Eq.~\eqref{eq:McLachlanVarQRTE} respectively Eq.~\eqref{eq:error_rqte_var}. Each evaluation is based on $10000$ samples that are measured from the quantum circuits. These quantities are then used to evaluate the ODE function to propagate the parameters and to compute the noisy error bound $\epsilon_{T}$. This error bound evaluation is compared to the Bures metric between the target state and the state underlying our noisy simulation given as a density matrix. 
The results are presented in Fig.~\ref{fig:ising_noisy_varqrte} for the residual as well as the gradient error ODE formulation. While the residual ODE formulation converges to the maximal value of $\sqrt{2}$ and, hence, does not enable to capture the true Bures metric, the dual formulation shows more robustness against the simulated noise, i.e., the final error bound is $0.4823$ while the actual error is $0.3018$.

\section{Conclusion and Outlook}
\label{sec:conclusions}
This work presents a posteriori error bounds for the Bures metric between a state prepared with VarQTE and the respective unkown target state resulting from exact QTE.
These bounds enable users to quantify the accuracy of their quantum time evolution simulation and potentially adapt their simulation setting if necessary.

The presented a posteriori, algorithmic error bounds for VarQTE lower bound existing error bounds \cite{MartinazzoErrorVarQuantumDyn20, Endo_2020VarQTEGeneralProcesses}. Furthermore, the bounds are particularly simple to evaluate, i.e., the additional resource overhead is limited to the evaluation of the energy variance.

We show that the error bounds and VarQTE itself are strongly dependent on the numerical integration method. An ODE solver which applies an adaptive step size scheme can increase the numerical stability and accuracy significantly. Furthermore, using an ODE formulation which is based on the minimization of the local gradient error $\|\ket{e_t}\|_2$ often helps to reduce the simulation errors.
The performance of the algorithm, the error bounds, related state fidelities, and system energies are demonstrated on numerical examples.

An open question for future research would be the investigation of the behavior of the error bounds at critical points, such as phase transitions. This study could give us important insights into the limits and potentials of the QTE simulation technique.
Furthermore, it would be of interest to conduct an enhanced study about the robustness of the error bounds under realistic quantum hardware conditions.

\vspace{3mm}

\noindent\textbf{Code Availability.}
The code can be made available upon reasonable request.
All quantities required to compute the presented error bounds can be evaluated with the tools provided by Qiskit's gradient framework:  \url{https://github.com/Qiskit/qiskit-terra/tree/main/qiskit/opflow/gradients}.

\noindent\textbf{Acknowledgments.}
We thank Pauline Ollitrault, Alexander Miessen, and Guglielmo Mazzola for insightful discussions on VarQRTE applications and Julien Gacon for proofreading this manuscript.
CZ acknowledges support from the National Centre of Competence in Research \emph{Quantum Science and Technology} (QSIT).

\appendix
\section{Inequality Relations of Distance Metrics}
\label{app:inequalities}
This section presents a formal introduction to the metric inequalities illustrated in Fig.~\ref{fig:inequalities}.
If the states are normalized, then the Bures metric simplifies to 
 \begin{align}
  \label{eq_buresPhase}
	B\left(\ket{\psi^{\omega}_t}, \ket{\psi^*_t} \right)&= \sqrt{2-2|\braket{\psi^{\omega}_t|\psi^*_t}|} \nonumber \\
	&=\!\min_{\phi \in [0,2\pi]}\!\norm{\ee^{i\phi}\ket{\psi^{\omega}_t} \!-\!\ket{\psi^*_t}}_2.
\end{align}
The last line highlights that the Bures metric can be interpreted as a global phase invariant $\ell_2$-norm and, hence, $B\left(\ket{\psi^{\omega}_t}, \ket{\psi^*_t} \right)\leq \norm{\ket{\psi^{\omega}_t} \!-\!\ket{\psi^*_t}}_2$.
The Bures metric is equivalent to the $\ell_2$-norm if
\begin{itemize}
    \item VarQTE does not induce a change in the global phase, or
    \item $\ket{\psi^{\omega}_t}$ can represent a global phase change, e.g., with an additional phase gate \cite{McKay_EfficientZ_2017}.
\end{itemize}
However, one may not a priori know whether a global phase change is induced by the considered Hamiltonian and an additional phase gate can introduce additional noise as well as imprecision in the parameter propagation.
Hence, the Bures metric offers an alternative to the $\ell_2$-norm which captures the properties of the systems while being agnostic to unphysical dependencies on global phases.

\begin{figure}[!ht]
    \centering
    \captionsetup{singlelinecheck = false, format= hang, justification=centerlast, font=footnotesize, labelsep=space}
    \begin{tikzpicture}
\node at (-1, 0.5) {\textbf{Trace Distance as a Function of the Bures Metric}};
\node[inner sep=0pt, anchor=north west] at (-4.5, -0.12) {\includegraphics[width=0.4\textwidth]{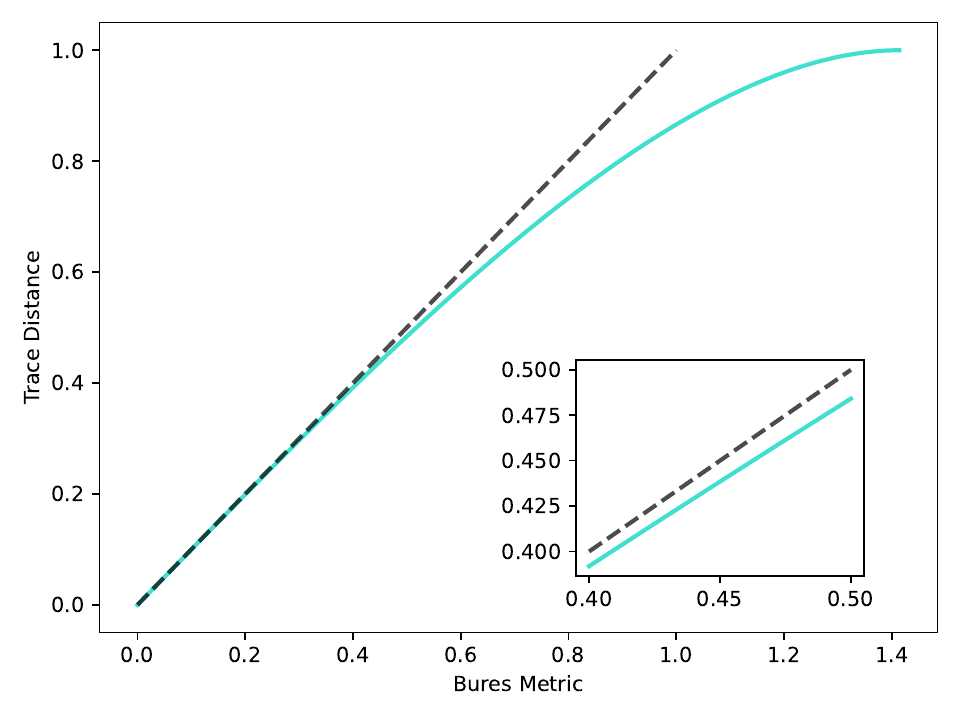}};
\end{tikzpicture}
    \caption{The figure shows the trace distance as a function of the Bures metric (solid line) and, hence, the trajectory on which an upper bound to the trace distance based on $\epsilon$ would exist. The dashed diagonal illustrates the Bures metric itself until the maximum trace distance value $1$ is reached.}
    \label{fig:trace_bures}
\end{figure}

Furthermore, the trace distance 
\begin{align}
    D_{\text{Tr}}\left(\ket{\psi^{\omega}_t}, \ket{\psi^*_t} \right) = \frac{1}{2} \norm{\ket{\psi^{\omega}_t}\bra{\psi^{\omega}_t}-\ket{\psi^*_t}\bra{\psi^*_t}},
\end{align}
where we use the simplified notation $D_{\text{Tr}}\left(\ket{\psi^{\omega}_t}, \ket{\psi^*_t} \right):=D_{\text{Tr}}\left(\proj{\psi^{\omega}_t}, \proj{\psi^*_t} \right)$, can be defined via the fidelity if the underlying states are pure
\begin{align}
    D_{\text{Tr}}\left(\ket{\psi^{\omega}_t}, \ket{\psi^*_t} \right) = \sqrt{1-|\braket{\psi^{\omega}_t|\psi^*_t}|^2}.
\end{align}
It follows that for pure states
\begin{align}
    D_{\text{Tr}}\left(\ket{\psi^{\omega}_t}, \ket{\psi^*_t} \right) = \sqrt{1-\left(1-\frac{B\left(\ket{\psi^{\omega}_t}, \ket{\psi^*_t} \right)^2}{2}\right)^2},
\end{align}
and, hence, that $ D_{\text{Tr}}\left(\ket{\psi^{\omega}_t}, \ket{\psi^*_t} \right)\leq B\left(\ket{\psi^{\omega}_t}, \ket{\psi^*_t} \right)$, where equality only holds if $\ket{\psi^{\omega}_t} = \ket{\psi^*_t}$. The relation is illustrated in Fig.~\ref{fig:trace_bures}.

\section{Global Phase Agnostic VarQRTE}
\label{app:global_phase_agnostic_varqrte}
We are now going to derive the ODE given in Eq.~\ref{eq:McLachlanVarQRTE}.
Consider the real time evolution of a parameterized state with an explicit time-dependent global phase parameter $\nu$, i.e., $\ket{\psi^{\nu}_t} = \ee^{-i\nu}\ket{\psi^{\omega}_t}$ for $\nu = \nu_t \in \mathbb{R}$,
where $
\ket{\dot\psi^\nu} = -i\dot\nu\ee^{-i\nu}\ket{\psi^{\omega}_t}+ \ee^{-i\nu}\ket{\dot{\psi}^{\omega}_t}$.
The Schr\"odinger equation with respect to $\ket{\psi^{\nu}_t}$ reads
\begin{align}
\label{eq:phase_fix_Schroed}
    i \ket{\dot\psi^\nu} = H\ket{\psi^\nu}
\end{align}
and can be rewritten as 
\begin{es}
    \label{eq:Schroedinger_real_nu}
     i \ee^{-i\nu}\ket{\dot\psi_t^{\omega}} = \left(H- \dot\nu\mathds{1}\right)\ee^{-i\nu}\ket{\psi_t^{\omega}}.
\end{es}
To simplify the notation, we shall from now on refer to $ \dot\nu\mathds{1}$ as $ \dot\nu$.
Division by $\ee^{-i\nu}$ gives
\begin{align}
\label{eq:redefinedVarQRTE}
     i \ket{\dot\psi^{\omega}_t} = \left(H- \dot\nu\right)\ket{\psi^{\omega}_t}.
\end{align}

Applying McLachlan's variational principle \cite{McLachlan64} to Eq.~\eqref{eq:redefinedVarQRTE} leads to
\begin{align}
\label{eq:MacLachlan_phase_real}
 \delta\norm{i \ket{\dot\psi^{\omega}_t} - \left(H- \dot\nu\right)\ket{\psi^{\omega}_t}}_2&= 0,
\end{align}
where $\norm{x}_2 = \sqrt{\langle x,x\rangle}$.
To find an explicit expression for $\dot\nu$, we evaluate the respective variational principle, i.e.,
\begin{align}
     \delta_{\dot\nu}\norm{i \ket{\dot\psi^{\omega}_t} - \left(H- \dot\nu\right)\ket{\psi^{\omega}_t}}_2&= 0
\end{align}
which leads to 
\begin{equation}
    \label{eq_nu_dot}
    \dot\nu = E_t^{\omega} + \text{Im}\left(\braket{\dot\psi^{\omega}_t|\psi^{\omega}_t}\right),
\end{equation}
where $E_t^{\omega}:=\bra{\psi^{\omega}_t}H\ket{\psi^{\omega}_t}$.

Finally, we can see that this formulation describes an evolution for $\ket{\psi^{\omega}_t}$ which simulates the existence of the global phase parameter $\nu$ without actually integrating or tracking $\ee^{-i\nu}$, i.e.,
\begin{align}
\label{eq:MacLachlan_phase_agnostic_real}
     i \ket{\dot\psi^{\dot\nu}_t}= H\ket{\psi^{\omega}_t},
\end{align}
where $\ket{\dot\psi^{\dot\nu}_t} := \ket{\dot\psi^{\omega}_t} -i(E_t^{\omega}+  \text{Im}(\braket{\dot\psi^{\omega}_t|\psi^{\omega}_t}))\ket{\psi^{\omega}_t}$ represents the effective state gradient.
McLachlan's variational principle now implies
\begin{es}
\label{eq:MacLachlan_mixed_real}
     \delta\| i \ket{\dot\psi^{\dot\nu}_t} - H\ket{\psi^{\omega}_t} \|_2&= 0.
\end{es}

Since $\ket{\psi^{\omega}_t}$ is given by a parameterized quantum circuit, solving Eq.~\eqref{eq:MacLachlan_mixed_real} with $\ket{\dot\psi^{\omega}_t} = \sum_i\dot\omega_i\frac{\partial\ket{\psi^{\omega}_t}}{\partial\omega_i}$
results in
 \begin{es}
      &\sum\limits_{j=0}^{k}\text{Re}\left(\frac{\partial \bra{\psi^{\omega}_t}}{\partial \omega_i}\frac{\partial \ket{\psi^{\omega}_t}}{\partial \omega_j} - \frac{\partial \bra{\psi^{\omega}_t}}{\partial \omega_i}\proj{\psi^{\omega}_t}\frac{\partial \ket{\psi^{\omega}_t}}{\partial \omega_j}\right) \dot\omega_j = \\
      &\text{Im}\left( \frac{\partial \bra{\psi^{\omega}_t}}{\partial \omega_i} H\ket{\psi^{\omega}_t} -  \frac{\partial \bra{\psi^{\omega}_t}}{\partial \omega_i}\ket{\psi^{\omega}_t}E_t^{\omega}\right).
\end{es}

\section{Global Phase Agnostic VarQITE}
\label{app:global_phase_agnostic_varqite}
Next, the derivation for the ODE given in Eq.~\ref{eq:McLachlanVarQITE} is presented.
The normalized, Wick-rotated Schr\"odinger equation of an evolution of the state $\ket{\psi^{\nu}_t} = \ee^{-i\nu}\ket{\psi^{\omega}_t}$ reads
\begin{es}
\label{eq:global_phase_wick}
         \ket{\dot\psi^\nu} = \left( E_t^{\omega} - H\right)\ket{\psi^\nu},
\end{es}
where
\begin{es}
     \ket{\dot\psi^\nu} = \ee^{-i\nu}\ket{\dot\psi^{\omega}_t} - i\dot\nu\ee^{-i\nu}\ket{\psi^{\omega}_t}.
\end{es}
Thus,
\begin{es}
     \ee^{-i\nu}\ket{\dot\psi^{\omega}_t} = \left(E_t^{\omega} - H +  i\dot\nu\right) \ee^{-i\nu}\ket{\psi^{\omega}_t}
\end{es}
and division by $\ee^{-i\nu}$ leads to
\begin{es}
     \ket{\dot\psi^{\omega}_t} = \left(E_t^{\omega} - H +  i\dot\nu\right)\ket{\psi^{\omega}_t}.
\end{es}
Application of McLachlan's variational principle gives
\begin{es}
     \delta\norm{\ket{\dot\psi^{\omega}_t} - \left(E_t^{\omega}-H+i\dot\nu\right)\ket{\psi^{\omega}_t}}_2&= 0.
\end{es}
Next, we evaluate the variational principle with respect to $\dot\nu$
\begin{es}
     \delta_{\dot\nu}\norm{\ket{\dot\psi^{\omega}_t} - \left(E_t^{\omega}-H+i\dot\nu\right)\ket{\psi^{\omega}_t} }_2 &= 0
\end{es}
and find that $\dot\nu = -\text{Im}(\braket{\dot\psi^{\omega}_t|\psi^{\omega}_t})$.
Now,
\begin{es}
     \label{eq:MacLachlan_phase_agnostic_imag}
     \ket{\dot\psi^{\dot\nu}_t}= \left(E_t^{\omega}-H\right)\ket{\psi^{\omega}_t}
\end{es}
simulates a global phase degree of freedom $\nu$ without actual implementation of $\ee^{-i\nu}$ and has the effective state gradient 
\begin{align} \label{eq_forDavid}
\ket{\dot\psi^{\dot\nu}_t} := \ket{\dot\psi^{\omega}_t} + i\text{Im}(\braket{\dot\psi^{\omega}_t|\psi^{\omega}_t})\ket{\psi^{\omega}_t}.
\end{align}
Rewriting the variational principle accordingly gives
\begin{es}
\label{eq:VarQITE_phase_agnostic}
     \delta\norm{\ket{\dot\psi^{\dot\nu}_t} - \Big(E_t^{\omega}-H\Big)\ket{\psi^{\omega}_t} }_2&= 0.
\end{es}

Since the time-dependence of $\ket{\psi^{\omega}_t}$ is encoded in the parameters $\bm{\omega}$, Eq.~\eqref{eq:VarQITE_phase_agnostic} leads to the following system of linear equations
\begin{es}
&\sum\limits_{j=0}^k \text{Re}\left(\frac{\partial \bra{\psi^{\omega}_t}}{\partial \omega_i}\frac{\partial \ket{\psi^{\omega}_t}}{\partial \omega_j} - \frac{\partial \bra{\psi^{\omega}_t}}{\partial \omega_i}\proj{\psi^{\omega}_t}\frac{\partial \ket{\psi^{\omega}_t}}{\partial \omega_j}\right)\dot\omega_j= \\
&- \text{Re}\left(\frac{\partial \bra{\psi^{\omega}_t}}{\partial \omega_i} H\ket{\psi^{\omega}_t}\right).
\end{es}

\section{Proof of Theorem~\ref{thm_VQRT}} \label{app:proofThm1}
For $\delta_t > 0$, let the state evolution be defined with respect to the effective gradient given in Eq.~\eqref{eq:MacLachlan_phase_agnostic_real}
    \begin{align} 
  \ket{\psi^{\omega}_{t+\delta_t}  }&= \ket{\psi^{\omega}_{t}  } + \delta_t\ket{\dot\psi^{\dot\nu}_t} \nonumber \\
  &=\ket{\psi^{\omega}_{t} } \!+\! \delta_t\!\left(\!\ket{\dot\psi^{\omega}_{t}}\!-\!iE_t^{\omega}  \!-\!i \text{Im}\left(\!\braket{\dot\psi^{\omega}_t|\psi^{\omega}_t}\!\right)\ket{\psi^{\omega}_t} \!\right)\!. \label{eq_step1}
    \end{align}
Combining Eq.~\eqref{eq_step1} with the triangle inequality gives 
\begin{align}
    B\left(\ket{\psi^{\omega}_{t+\delta_t}}, \ket{\psi^*_{t+\delta_t}}\right) 
      &\leq B\Big(\ket{\psi^{\omega}_{t+\delta_t}}, \big(\mathds{1}\!-\!i\delta_tH\big)\ket{\psi^{\omega}_{t}}\! \Big)\nonumber  \\
      &\hspace{-6mm}+B\Big(\big(\mathds{1}-i\delta_tH\big)\ket{\psi^{\omega}_{t}}, \ket{\psi^*_{t+\delta_t}}\Big). \label{eq_triangle_start}
\end{align}
Using Eq.~\eqref{eq_buresPhase} and neglecting terms of order $\mathcal{O}(\delta_t^2)$ gives
\begin{align}
     &B\Big(\ket{\psi^{\omega}_{t+\delta_t}}, \big(\mathds{1}-i\delta_tH\big)\ket{\psi^{\omega}_{t}} \Big) \nonumber \\
     &\hspace{5mm}=\min_{\phi \in [0,2\pi]}\norm{\ee^{i\phi}\big(\ket{\psi^{\omega}_{t+\delta_t}} \big) - \big(\mathds{1}-i\delta_tH\big)\ket{\psi _{t}}}_2 \nonumber\\
     &\hspace{5mm}\leq \norm{\ket{\psi^{\omega}_{t+\delta_t}}- \big(\mathds{1}-i\delta_tH\big)\ket{\psi _{t}}}_2 \nonumber\\
     &\hspace{5mm}=\delta_t \norm{\ket{\dot\psi^{\omega}_t}  + i\big(H\!-\!E_t^{\omega}\!-\! \,\text{Im}(\braket{\dot\psi^{\omega}_t|\psi^{\omega}_t}) \big)\ket{\psi^{\omega}_t}
    }_2 \nonumber \\
     &\hspace{5mm}=: \delta_t\norm{\ket{e_t}}_2, \label{eq_triangle_part1}
\end{align}
where the penultimate step uses Eq.~\eqref{eq_nu_dot}.

For the second term in Eq.~\eqref{eq_triangle_start}, we employ a simplified notation for QRTE for a time step $\delta t$, i.e,
\begin{align}
    \mathcal{P}^{\text{real}}_{\delta t}\big(\ket{\phi_t}\big) := \ket{\phi_{t+\delta t}} = \big(\mathds{1} - iH\big) \ket{\phi_t},
\end{align}
which leads to 
\begin{align}
\label{eq_triangle_part2}
     &B\Big(\big(\mathds{1}-i\delta_tH\big)\ket{\psi^{\omega}_{t}}, \ket{\psi^*_{t+\delta_t}}\Big) \\
     &\hspace{0mm}=B\Big(\!\mathcal{P}^{\text{real}}_{\delta t}\big(\ket{\psi^{\omega}_t}\big), \mathcal{P}^{\text{real}}_{\delta t}\big(\ket{\psi^*_{t+\delta_t}}\big) \Big) \\
     &\leq B\Big(\!\ket{\psi^{\omega}_t}, \ket{\psi^*_{t+\delta_t}} \Big),
\end{align}
The penultimate step uses that all physical processes are non-trace-increasing \cite{nielsen10} which implies that the Bures metric does not increase either.

Combining Eqs~\eqref{eq_triangle_start},~\eqref{eq_triangle_part1} and~\eqref{eq_triangle_part2} gives
\begin{align}
     B\left(\ket{\psi^{\omega}_{t+\delta_t}}, \ket{\psi^*_{t+\delta_t}}\right)
     \leq B\left(\ket{\psi^{\omega}_t}, \ket{\psi_t^*}\right) + \delta_t \norm{\ket{e_t}}_2.
\end{align}
Assuming that $B\left(\ket{\psi_0}, \ket{\psi_0^*}\right) = 0$, we can evolve
\begin{align}
B\left(\ket{\psi^{\omega}_{T}}, \ket{\psi_{T}^*}\right) = \delta_t\sum\limits_{k=0}^{K}\|\ket{e_{k\delta_t}}\|_2,
\end{align}
where $K$ corresponds to the number of time steps.
Finally setting $\delta_t = T/t$ leads to
\begin{align}
     B\left(\ket{\psi^{\omega}_{T}}, \ket{\psi_{T}^*}\right) \leq \int_{0}^{T} \|\ket{e_t}\|_2 \, \di t := \epsilon_T,
\end{align}
which proves the assertion.\qed

\section{Proof of Theorem~\ref{thm_VQITE}} \label{app:proofThm2}
Combining that Eq.~\eqref{eq:MacLachlan_phase_agnostic_imag} gives
    \begin{align} 
  \ket{\psi^{\omega}_{t+\delta_t}  }&= \ket{\psi^{\omega}_t  } + \delta_t\ket{\dot\psi^{\dot\nu}_t} \nonumber   \\
  &=\ket{\psi^{\omega}_t  } + \delta_t\left(\ket{\dot\psi^{\omega}_t} + i\text{Im}(\braket{\dot\psi^{\omega}_t|\psi^{\omega}_t})\ket{\psi^{\omega}_t}\right),
    \end{align}
 with the triangle inequality for $\delta_t > 0$ results in. 
\begin{align}
    &B\left(\ket{\psi^{\omega}_{t+\delta_t}}, \ket{\psi^*_{t+\delta_t}}\right) \nonumber \\ 
      &\hspace{0mm}\leq B\Big(\!\ket{\psi^{\omega}_t  } \!+\! \delta_t\ket{\dot\psi^{\dot\nu}_{t}}, \big(\mathds{1}\!+\delta_t\left(E_t^{\omega}-H\right)\big)\ket{\psi^{\omega}_t}\! \Big)\nonumber  \\
      &\hspace{2mm}+B\Big(\!\big(\mathds{1}\!+\delta_t\left(E_t^{\omega}-H\right)\big)\ket{\psi^{\omega}_t}, \ket{\psi^*_{t+\delta_t}}\Big) . \label{eq_triangle_start_qite}
\end{align}
Next, we consider the two terms separately.
Using Eq.~\eqref{eq_buresPhase} and neglecting terms of order $\mathcal{O}(\delta_t^2)$ gives
\begin{align}
     &B\Big(\ket{\psi^{\omega}_t  } \!+\! \delta_t\ket{\dot\psi^{\dot\nu}_{t}}, \big(\mathds{1}\!+\delta_t\left(E_t^{\omega}-H\right)\big)\ket{\psi^{\omega}_t} \Big) \nonumber \\
     &=\min_{\phi \in [0,2\pi]}\norm{\ee^{i\phi}(\ket{\psi^{\omega}_t  } \!+\! \delta_t\ket{\dot\psi^{\dot\nu}_{t}}) - \big(\mathds{1}\!+\delta_t\left(E_t^{\omega}-H\right)\big)\ket{\psi^{\omega}_t}}_2 \nonumber\\
     &\leq\norm{\ket{\psi^{\omega}_t} + \delta_t\ket{\dot\psi^{\dot\nu}_{t}}  - \left(\mathds{1}+\delta_t\left(E_t^{\omega}-H\right)\right)\ket{\psi^{\omega}_t}}_2  \nonumber\\
     &= \delta_t \norm{\ket{\dot\psi^{\dot\nu}_{t}}  - \left(E_t^{\omega}-H\right)\ket{\psi^{\omega}_t}}_2  \nonumber\\
     &=\delta_t \norm{\ket{\dot\psi^{\omega}_t}  - \left(E_t^{\omega}-H-i\text{Im}\left(\braket{\dot\psi^{\omega}_t|\psi^{\omega}_t}\right)\right)\ket{\psi^{\omega}_t}
    }_2 \nonumber \\
     &= \delta_t\norm{\ket{e_t}}_2, \label{eq_triangle_part1_qite}
\end{align}
where the penultimate step uses Eq.~\eqref{eq_forDavid}.

For the second term in Eq.~\eqref{eq_triangle_start_qite}, we employ a simplified notation for exact QITE for a time step $\delta t$, i.e,
\begin{align}
    \mathcal{P}^{\text{imag}}_{\delta t}\big(\ket{\phi_t}\big) := \ket{\phi_{t+\delta t}} = \big(\mathds{1}\bra{\phi_t}H\ket{\phi_t} + H\big) \ket{\phi_t},
\end{align}
which leads to 
\begin{align}
\label{eq_triangle_part2_qite}
     &B\Big(\!\big(\mathds{1}\!+\delta_t\left(E_t^{\omega}-H\right)\big)\ket{\psi^{\omega}_t}, \ket{\psi^*_{t+\delta_t}}\Big) \\
     &\hspace{0mm}=B\Big(\!\mathcal{P}^{\text{imag}}_{\delta t}\big(\ket{\psi^{\omega}_t}\big), \mathcal{P}^{\text{imag}}_{\delta t}\big(\ket{\psi^*_{t+\delta_t}}\big) \Big) \\
     &\leq B\Big(\!\ket{\psi^{\omega}_t}, \ket{\psi^*_{t+\delta_t}} \Big),
\end{align}
where the last line holds because all physical processes are non-trace-increasing \cite{nielsen10}. 

Combining Eqs~\eqref{eq_triangle_start_qite},~\eqref{eq_triangle_part1_qite} and~\eqref{eq_triangle_part1_qite} gives
\begin{align}
     B\left(\ket{\psi^{\omega}_{t+\delta_t}}, \ket{\psi^*_{t+\delta_t}}\right)
     \leq B\left(\ket{\psi^{\omega}_t}, \ket{\psi_t^*}\right) + \delta_t \norm{\ket{e_t}}_2.
\end{align}
The final steps of the proof are equivalent to the ones presented in Appendix \ref{app:proofThm1}.

\qed

\section{VarQTE Implementation}
\label{app:implementation}

The implementation of VarQTE relies on the evaluation of
$\text{Im}(C_i - \textstyle{\frac{\partial \bra{\psi^{\omega}_t}}{\partial \omega_i}\ket{\psi^{\omega}_t}E_t^{\omega}})$, $\text{Re}(C_i)$ and $
\mathcal{F}^Q_{ij}$ which are introduced in Sec.~\ref{sec:VarQTE}.
The parameterized state is constructed as $\ket{\psi^{\omega}_t} = \prod_{p=0}^k U_p(\omega_p)\ket{0}^{\otimes n}$. Thus, we may use that parameterized unitaries can be written as $\textstyle{U_j\left(\omega_j\right) = \ee^{iM\left(\omega_j\right)}}, $
where $M\left(\omega_j\right)$ denotes a parameterized Hermitian matrix. 
To simplify the notation, we assume that $M(\omega_j)=\textstyle{ {-\frac{\omega_j}{2}\sigma_j}}$ for $\sigma_j\in\set{\mathds{1}, X, Y, Z}$.
Since
\begin{equation}
\frac{\partial U_j\left(\omega_j\right)}{\partial\omega_j} = -\frac{i}{2} \sigma_jU_j\left(\omega_j\right),
\end{equation}
it follows that
\begin{align}
\label{eq:gradient}
\frac{\partial\ket{\psi^{\omega}_t}}{\partial\omega_j} = -\frac{i}{2}\prod\limits_{p=j+1}^k U_p\left(\omega_p\right) \sigma_jU_j\left(\omega_j\right)\prod\limits_{p=0}^{j-1} U_p\left(\omega_p\right)\ket{0}^{\otimes n}.
\end{align}
Next, we employ Eq.~\eqref{eq:gradient} to find that
\begin{align}
   C_i &= -\frac{i}{2}\bra{\psi^{\omega}_t} H\prod\limits_{p=j+1}^k U_p \sigma_jU_j\prod\limits_{p=0}^{j-1} U_p\ket{0}^{\otimes n},
\end{align}
as well as,
\begin{align}
    &\frac{\partial \bra{\psi^{\omega}_t}}{\partial \omega_i}\ket{\psi^{\omega}_t}
=  \frac{i}{2}\bra{0}^{\otimes n}\prod\limits_{p=0}^{j-1} U^{\dagger}_p U^{\dagger}_j\sigma_j\prod\limits_{p=j+1}^k U^{\dagger}_p\ket{\psi^{\omega}_t},
\end{align}
and
\begin{align}
\mathcal{F}^Q_{ij} &= \frac{1}{4}
\text{Re}\Big(\bra{0}^{\otimes n} \prod\limits_{p=0}^{i-1} U^{\dagger}_pU^{\dagger}_i\sigma_i\prod\limits_{p=i+1}^{j-1} U^{\dagger}_p
 \sigma_j\prod\limits_{p=0}^{j-1} U_p\ket{0}^{\otimes n}\nonumber\\
&\hspace{5mm} -\bra{0}^{\otimes n} \prod\limits_{p=0}^{i-1} U^{\dagger}_pU^{\dagger}_i\sigma_i\prod\limits_{p=i+1}^k U^{\dagger}_p \proj{\psi^{\omega}_t}
\nonumber\\
&\hspace{5mm}\prod\limits_{p=j+1}^k U_p \sigma_jU_j\left(\omega_j\right)\prod\limits_{p=0}^{j-1} U_p\ket{0}^{\otimes n}
\Big),
\end{align}
where we assume that $i<j$ and simplify the notation with $U_p := U_p\left(\omega_p\right)$.

\begin{figure}[!htb]
\captionsetup{singlelinecheck = false, format= hang, justification=centerlast, font=footnotesize, labelsep=space}
\begin{center}
\begin{tikzpicture}
\node at (0,0) {\includegraphics[width=0.70\linewidth]{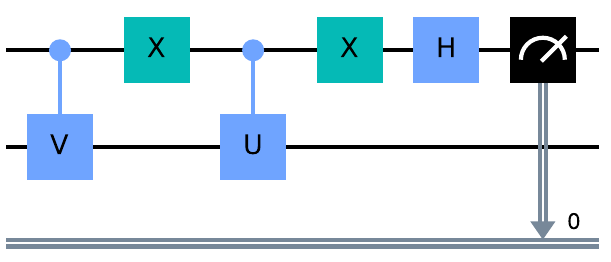}};
\node at (-4.0,0.8) {$\ket{0}+e^{i \alpha} \ket{1}$};
\node at (-3.5,-0.15) {$\ket{\psi_{\textnormal{in}}}$};
\node at (-3.3,-1.1) {$c$};
\node at (2.4,1.35) {$Z$};
\end{tikzpicture}
\end{center}
\caption{This quantum circuit -- originally proposed in \cite{LaflammeSimulatingPhysPhenom02} -- uses an additional working qubit to evaluate $\text{Re}\left(\ee^{i\alpha}\bra{\psi_{\text{in}}}U^{\dagger}V\ket{\psi_{\text{in}}}\right)$. Notably, this only requires to measure the working qubit with respect to $Z$.}
\label{fig:expValueCircuit}
\end{figure}
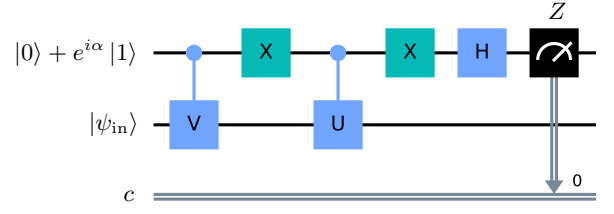

\begin{figure}[!htb]
\captionsetup{singlelinecheck = false, format= hang, justification=centerlast, font=footnotesize, labelsep=space}
\begin{center}
\begin{tikzpicture}
\node at (-3.2,1.25) {$\ket{0}+e^{i \alpha} \ket{1}$};
\node at (-2.7,0.0) {$\ket{\psi_{\textnormal{in}}}$};
\node at (-2.5,-1.35) {$c$};
\node at (1.4,2) {$Z$};
\node at (0,0) {\includegraphics[width=0.5\linewidth]{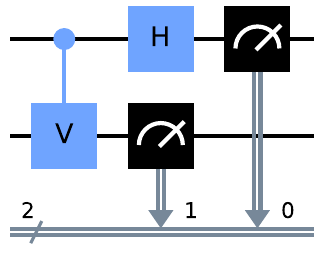}};
\node at (0.05,0.6) {$H$};
\end{tikzpicture}
\end{center}
\caption{This quantum circuit uses an additional working qubit to evaluate $\text{Re}(\ee^{i\alpha}\bra{\psi_{\text{in}}}HV\ket{\psi_{\text{in}}})$, where the working qubit is measured with respect to $Z$ and the state $\ket{\psi_{\text{in}}}$ with respect to the observable \smash{$H$}.}
\label{fig:expValueCircuitwithH}
\end{figure}
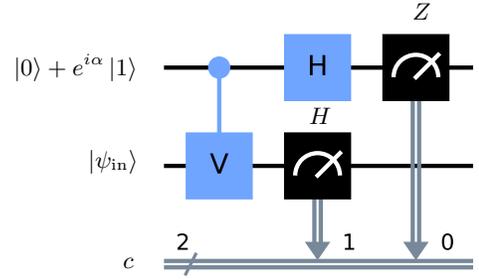

One can, now, see that
$\text{Im}(C_i -  \textstyle{\frac{\partial \bra{\psi^{\omega}_t}}{\partial \omega_i}\ket{\psi^{\omega}_t}E_t^{\omega})}$, $\text{Re}(C_i)$ and $
\mathcal{F}^Q_{ij}$ may be decomposed into terms of the form $
    \text{Re}(\ee^{i\alpha}\bra{\psi_{\text{in}}}UV\ket{\psi_{\text{in}}})$, respectively $    \text{Re}(\ee^{i\alpha}\bra{\psi_{\text{in}}}HV\ket{\psi_{\text{in}}})$
using that $\text{Im}(iz)=\text{Re}(z)$.
We can, thus, evaluate the equations either with the quantum circuit shown in Fig.~\ref{fig:expValueCircuit} or the one presented in Fig.~\ref{fig:expValueCircuitwithH}.

Consider, for example, $\ket{\psi^{\omega}_t} = \textstyle{\ee^{-i\frac{\omega_1}{2}X}\ee^{-i\frac{\omega_0}{2}Y}\ket{0}}$. 
Then,
\begin{align}
\frac{\partial \bra{\psi^{\omega}_t}}{\partial \omega_1}\ket{\psi^{\omega}_t} = \frac{i}{2}\bra{\psi^{\omega}_t} X\ket{\psi^{\omega}_t}.
\end{align}
and
\begin{align}
    C_1= -\frac{i}{2}\bra{\psi^{\omega}_t} HX\ket{\psi^{\omega}_t}.
\end{align}
To evaluate $\text{Im}(\textstyle{\frac{\partial \bra{\psi^{\omega}_t}}{\partial \omega_1}\ket{\psi^{\omega}_t}})$ with the circuit shown in Fig.~\ref{fig:expValueCircuit}, we set $\alpha=-\pi$, $\ket{\psi_{\text{in}}}=\ket{\psi^{\omega}_t}$, $V=X$ and $U=\mathds{1}$. Furthermore, $\text{Im}\left(C_1\right)$, respectively $\text{Re}\left(C_1\right)$ can be computed using Fig.~\ref{fig:expValueCircuitwithH} with $\alpha=0$, respectively $\alpha=-\pi/2$, $\ket{\psi_{\text{in}}}=\ket{\psi^{\omega}_t}$ and $V=X$.
Similarly, the evaluation of
\begin{align}
    \mathcal{F}^Q_{01} &= \frac{1}{4}
\text{Re}\Big(\bra{0}\ee^{i\frac{\omega_0}{2}Y}Y X\ee^{-i\frac{\omega_0}{2}Y}\ket{0} \nonumber \\
&\hspace{5mm}- \bra{0}Y\ket{0}\bra{0}\ee^{i\frac{\omega_0}{2}Y}X\ee^{-i\frac{\omega_0}{2}Y}\ket{0}\Big)\nonumber \\
&=\text{Re}\Big(\bra{0}\ee^{i\frac{\omega_0}{2}Y}Y X\ee^{-i\frac{\omega_0}{2}Y}\ket{0}\Big).
\end{align}
may be conducted with the setup illustrated in Fig.~\ref{fig:expValueCircuit} using $\alpha=0$, $\ket{\psi_{\text{in}}} = \textstyle{\ee^{-i\frac{\omega_0}{2}Y}\ket{0}}$, $V=X$ and $U=Y$.

\section{Larger Ising Model Simulations}
\label{app:10qubits}
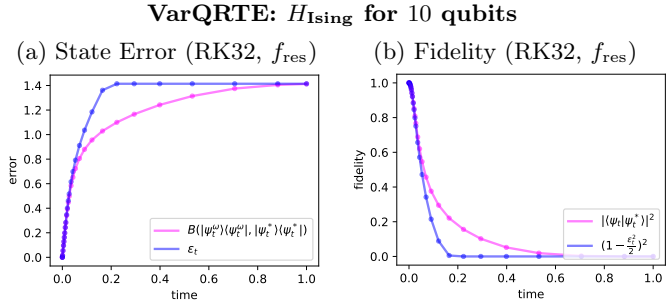
\begin{figure}[]
    \centering
    \captionsetup{singlelinecheck = false, format= hang, justification=centerlast, font=footnotesize, labelsep=space}
    \begin{tikzpicture}
\node at (-1, 0.5) {\textbf{VarQRTE: $H_{\text{Ising}}$ for $10$ qubits}};
\node[inner sep=0pt, anchor=north west] at (-5.5, -0.12) {\includegraphics[width=0.25\textwidth]{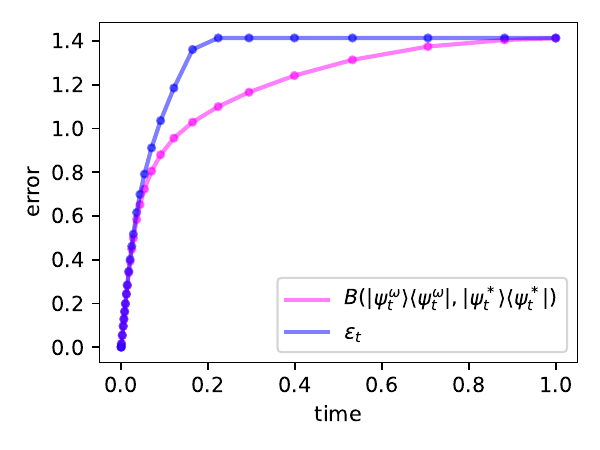}};
\node[anchor=north west] at (-1,0) {    \includegraphics[width=0.25\textwidth]{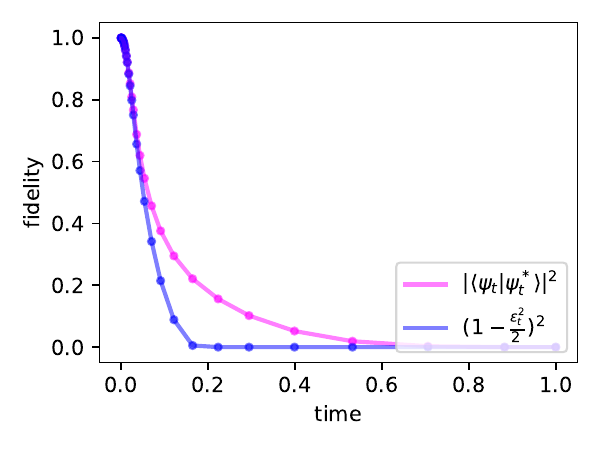}};
\node at (-3.2, 0) {(a) State Error (RK32, $ f_{\text{res}}$)};
\node at (1.3, 0) {(b) Fidelity (RK32, $ f_{\text{res}}$)};
\end{tikzpicture}
    \caption{VarQRTE for $\ket{\psi_0}=\ee^{-i\alpha}\ket{0}^{\otimes 10}$, $H_{\text{Ising}}$ and $T=1$ with RK32 is based on the residual ODE. (a) shows the error bounds $\epsilon_t$ and the actual Bures metric. (b) illustrates the corresponding fidelity and fidelity bound.}
    \label{fig:ising_10_varqrte}
\end{figure}

\begin{figure}[!ht]
    \centering
    \captionsetup{singlelinecheck = false, format= hang, justification=centerlast, font=footnotesize, labelsep=space}
    \begin{tikzpicture}
\node at (-1, 0.5) {\textbf{VarQITE: $H_{\text{Ising}}$ for $10$ qubits}};
\node[inner sep=0pt, anchor=north west] at (-5.5, -0.12) {\includegraphics[width=0.25\textwidth]{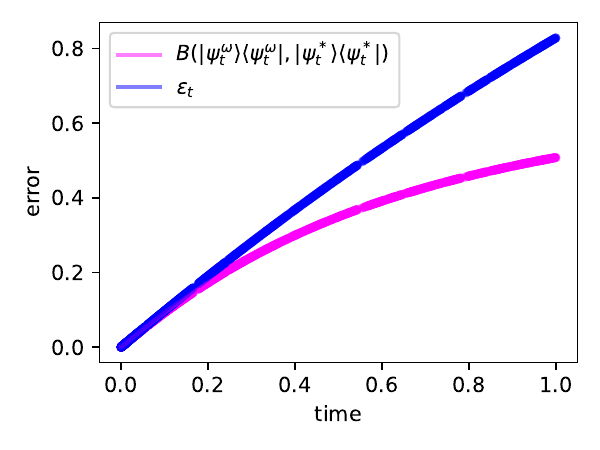}};
\node[anchor=north west] at (-1,0) {    \includegraphics[width=0.25\textwidth]{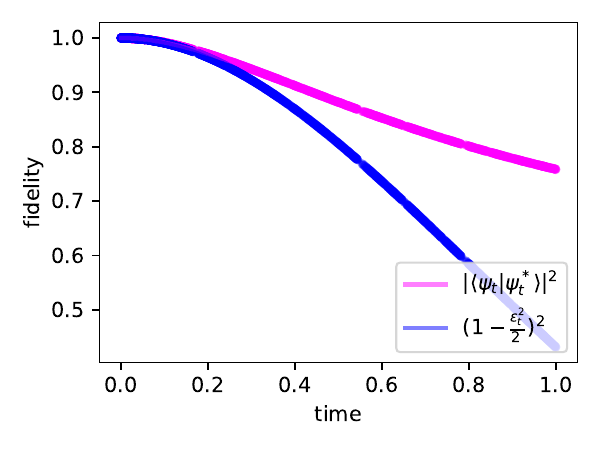}};
\node at (-3.2, 0) {(a) State Error (RK32, $ f_{\text{err}}$)};
\node at (1.3, 0) {(b) Fidelity (RK32, $ f_{\text{err}}$)};
\end{tikzpicture}
    \caption{VarQITE for $\ket{\psi_0}=\ee^{-i\alpha}\ket{0}^{\otimes 10}$, $H_{\text{Ising}}$ and $T=1$ with RK32 is based on the error based ODE. (a) shows the error bounds $\epsilon_t$ and the actual Bures metric. (b) illustrates the corresponding fidelity and fidelity bound.}
    \label{fig:ising_10_varqite}
\end{figure}

In order to rule out that the applicability of the presented error bounds is limited to systems consisting only of a few qubits, one has to conduct further experiments for larger system dimensions.
To progress towards this understanding, we extend our experimental analysis to the Ising model described in Sec.~\ref{sec:results} with $10$ qubits.
More specifically, we run VarQRTE and VarQITE simulations for $T=1$ using a Runge-Kutta method of order 3(2) (RK32) from SciPy~\cite{2020SciPy-NMeth}.
Fig.~\ref{fig:ising_10_varqrte} shows the results of the VarQRTE experiment that employs the residual ODE and ridge regression to solve the underlying SLE. We can see that the error bound matches the actual error up to a factor $10^{-1}$ until the Bures distance reaches approximately $0.8$. This directly relates to a fidelity of $0.68$.
In the VarQITE setup, the propagation is based on the error based ODE and the respective SLE is solved with a least squares approach. The results are presented in Fig.~\ref{fig:ising_10_varqite}.
The plots show that the ODE solver requires many time steps indicating a volatile propagation. The resulting error bounds then diverge by more than a factor $10^{-1}$ when the Bures distance reaches $0.35$ which corresponds to a fidelity of $0.94$.

\bibliography{references}

\end{document}